\documentclass[useAMS,usenatbib]{mn2e}
\usepackage{epsf,rotating}
\usepackage{amsmath}
\usepackage{subfigure}

\newcommand{\kmsmpc}{\kms\;{\rm Mpc}^{-1}}
\newcommand{\lya}{Ly$\alpha$}
\newcommand{\nhi}{N_{\rm HI}}
\newcommand{\hkpc}{h^{-1}{\rm kpc}}
\newcommand{\hmpc}{h^{-1}{\rm Mpc}}

\newcommand{\ghi}{\Gamma_{\rm HI}}
\newcommand{\kms}{\;{\rm km}\,{\rm s}^{-1}}
\newcommand\cdunits{{\rm cm}^{-2}}

\newcommand{\gad}{{\sc Gadget-2}}
\newcommand{\autovp}{{\sc AutoVP}}
\newcommand{\ion}[2]{\hbox{#1\,{\sc #2}}}
\newcommand{\vw}{v_{\rm w}}
\newcommand{\cm}{\,{\rm cm}}

\title[The Low-$z$ IGM]{The intergalactic medium over the last 10 billion
years I: Lyman alpha absorption and physical conditions}

\author[R. Dav\'e et al.]{
\parbox[t]{\textwidth}{\vspace{-1cm}
Romeel Dav\'e$^1$, Benjamin D. Oppenheimer$^2$, Neal Katz$^3$, 
Juna A. Kollmeier$^4$, David H. Weinberg$^5$}
\\\\$^1$ Astronomy Department, University of Arizona, Tucson, AZ 85721, USA
\\$^2$ Leiden Observatory, Leiden University, PO Box 9513, 2300 RA Leiden, Netherlands
\\$^3$ Astronomy Department, University of Massachussetts, Amherst, MA 01003, USA
\\$^4$ Observatories of the Carnegie Institute of Washington, Pasadena, CA 91101, USA
\\$^5$ Astronomy Department and CCAPP, 
Ohio State University, Columbus, OH 43210, USA
}

\begin{document}

\pubyear{2010}

\maketitle

\label{firstpage}

 \begin{abstract}
The intergalactic medium (IGM) is the dominant reservoir of baryons at all
cosmic epochs.  In this paper, we investigate the evolution of the IGM from
$z=2\rightarrow 0$ in $(48\hmpc)^3$, 110-million particle cosmological
hydrodynamic simulations using three prescriptions for galactic outflows.
We focus on the evolution of IGM physical properties, and how such
properties are traced by \lya\ absorption as detectable using {\it
Hubble's} Cosmic Origins Spectrograph (COS).  Our results broadly confirm
the canonical picture that most \lya\ absorbers arise from highly ionized
gas tracing filamentary large-scale structure.  Growth of structure causes
gas to move from the diffuse photoionized IGM into other cosmic phases,
namely stars, cold and hot gas within galaxy halos, and the unbound and
shock-heated warm-hot intergalactic medium (WHIM).  By today, baryons
are comparably divided between bound phases (35\% in our favoured
outflow model), the diffuse IGM (41\%), and the WHIM (24\%).  Here we
(re)define the WHIM as gas with overdensities lower than that in halos
($\rho/\bar{\rho}\la 100$ today) and temperatures $T>10^{5}$~K, 
to more closely align it with the ``missing baryons" that are not easily
detectable in emission or \lya\ absorption.  Strong galactic outflows can
have a noticeable impact on the temperature of the IGM, though with our
favoured momentum-driven wind scalings they do not.  When we (mildly) tune
our assumed photoionizing background to match the observed evolution of
the \lya\ mean flux decrement, we obtain line count evolution statistics
that broadly agree with available (pre-COS) observations.  We predict
a column density distribution slope of $f(\nhi)\propto \nhi^{-1.70}$
for our favoured wind model, in agreement with recent observational
estimates, and it becomes shallower with redshift.  Winds have a mostly
minimal impact, but they do result in a shallower column density slope
and more strong lines.  With improved statistics, the frequency of strong
lines can be a valuable diagnostic of outflows, and the momentum-driven
wind model matches existing data significantly better than the two
alternatives we consider.  The relationship between column density
and physical density broadens mildly from $z=2\rightarrow 0$, and evolves as
$\rho\propto \nhi^{0.74} 10^{-0.37z}$ for diffuse absorbers,
consistent with previous studies.  Linewidth distributions are
quite sensitive to spectral resolution; COS should yield significantly
broader lines than higher-resolution data.  Thermal contributions
to linewidths are typically subdominant, so linewidths only loosely
reflect the temperature of the absorbing gas. This will hamper attempts
to quantify the WHIM using broad \lya\ absorbers, though it may still
be possible to do so statistically.  Together, COS data and simulations
such as these will provide key insights into the physical conditions of
the dominant reservoir of baryons over the majority of cosmic time.
\end{abstract}

\begin{keywords}
galaxies: formation, large-scale structure, quasars: absorption lines, ultraviolet: general
\end{keywords}
 
\section{Introduction}

Diffuse intergalactic hydrogen produces numerous weak redshifted
\ion{H}{i} \lya~(1216\AA) absorption features along lines of sight
to distant bright objects such as quasars, the phenomenon known as the
\lya\ forest~\citep{lyn71,sar80}.  Advances in sensitive high-resolution
spectroscopy and theoretical understanding of structure
formation have led to the currently favoured paradigm for the origin
of the \lya\ forest~\citep{cen94,zha95,mir96,her96}: it arises from
highly photoionized intergalactic hydrogen~\citep{gun65} tracing uncollapsed
large-scale structure in the Cosmic Web.  The general properties
of the \lya\ forest are well-described by the Fluctuating Gunn-Peterson
Approximation~\citep[FGPA;][]{wei97a,cro98}, 
in which a
tight relation between density and temperature in cosmologically-expanding
photoionized gas~\citep{hui97} leads to a tight correlation between
the \lya\ optical depth and the underlying mass density at each location
along the line of sight.
At $z\ga 2$, where the \lya\
transition falls into the observed-frame
optical, this approximation is quite good,
and it can be used to infer the 
mean baryon density \citep{rau97,wei97b} and the
matter fluctuation spectrum, which in turn
provides constraints on cosmological 
parameters \citep{cro99,cro02,mcd00,mcd06,vie04,vie08}.

In this paper, a successor to \cite{dav99} and \cite{dav01},
we use cosmological hydrodynamic simulations to model the IGM
and the \lya\ forest at lower redshifts.  Here
the \lya\ transition falls in the ultraviolet
(UV), requiring more difficult space-based observations.  Furthermore,
the physical description 
of the \lya\ forest departs more strongly from the FGPA,
because filamentary structures begin to grow so large as to heat a
significant fraction of intergalactic gas to well above photoionization
temperatures \citep{dav99}.  This results in the emergence of
the Warm-Hot Intergalactic Medium (WHIM; 
\citealt{cen99,dav99,dav01,cen06}), which contains
a substantial fraction of all baryons today.  Additionally, the evolving
relationship between matter overdensity and \lya\ column density
is such that at low-$z$, a given overdensity produces a lower column
density absorber:  
roughly speaking, a $10^{14.5}\cm^{-2}$ line at $z\approx 2.5$ is
physically analogous to a $10^{13}\cm^{-2}$ line at $z=0$ \citep{dav99}.
This shift results in many fewer \lya\ forest
absorbers to a given sensitivity limit, i.e. a much thinner forest of
lines.
All these factors make the low-$z$ \lya\ forest more difficult to
characterize both observationally and theoretically, and not as useful for
constraining cosmology~\citep{zha05}.  But the intergalactic medium (IGM)
at low-$z$ still contains the vast majority of baryons in the Universe,
so understanding the low-$z$ \lya\ forest and the WHIM remains a central
challenge to assembling a complete picture of cosmic baryon evolution.

One of the original Key Projects of {\it Hubble Space Telescope} 
was to understand
the low-$z$ \lya\ forest.  The Faint Object Spectrograph (FOS), despite
being unable to fully resolve \lya\ absorbers, provided critical insights
into the nature of the low-$z$ \lya\ forest.  For example, even the earliest
observations showed
that the number density evolution of \lya\ absorbers must slow dramatically
between $z>2$ and $z=0$ \citep{bah91,mor91}.  Initially, it was suggested that
low-$z$ absorbers might represent a different population than those at
high-$z$, with the high-$z$ systems tracing the Cosmic Web and the low-$z$
systems arising in galaxy halos~\citep{bah96}.  This idea was supported
by the apparently tight connection between column density and galaxy
impact parameter~\citep{che98}.  However, \citet{dav99} showed that
all these observations can be fully accomodated within the structure
formation model of the \lya\ forest.  The change in number density
evolution can be explained as an ionization effect~\citep{the98}, since
the dominant source of metagalactic flux (quasars) diminishes rapidly
at $z\la 2$~\citep{haa01},  
countering the effect of the declining mean density of baryons.
The relationship between \lya\ equivalent
width and galaxy impact parameter arises because the IGM is denser around
galaxies owing to matter clustering.  Subsequent higher-resolution quasar
spectroscopy using the Space Telescope Imaging Spectrograph (STIS)
has generally confirmed these interpretations, and STIS \lya\ forest
data is well-described by structure formation models~\citep{dav01b}.
Hence the low-$z$ \lya\ forest is not fundamentally different than the
high-$z$ one, although the relationship of neutral hydrogen absorption
to large-scale structure has shifted.

The installation of the Cosmic Origins Spectrograph (COS) on {\it
Hubble} should usher in a new age for the study of the
low-$z$ \lya\ forest.  The dramatically increased sensitivity over
STIS will enable a more robust characterization of the $z\la 1$
IGM, and it will enable study of \lya\ absorbers at overdensities comparable
to that detectable at high-$z$.  COS will also enable tomographic
probes of the IGM at transverse separations that will provide
interesting constraints on the large-scale coherence of \lya\
absorbers~\citep[e.g.][]{cas08}.  All these data will provide
theoretical studies with new opportunities to understand the low-$z$
IGM, and new challenges for models to meet.

An advantage of low-$z$ IGM work compared to high-$z$ is that the galaxy
population around low-$z$ absorbers can be studied to much greater
depth.  The relation between the two provides unique insights into how
the early and late stages of gravitational collapse are connected,
and how the baryons have decoupled from the dark matter via shock
heating and radiative cooling.  The galaxy-absorber connection can be
studied through the individual association of galaxies with absorbers
or by statistical measures such as correlation functions.  This is
the subject of a subsequent paper in this series (Kollmeier et al.,
in preparation).  Another paper in this series (Oppenheimer et al., in
preparation) will deal with metal line absorbers detectable using COS,
and their relationship to galaxies and large-scale structure.

In this paper, we employ state-of-the-art cosmological hydrodynamic
simulations to study the $z\la 1$ IGM as traced by the \lya\ forest
and WHIM.  Where appropriate we include comparisons with existing
data and predictions for COS.  Our new simulations have $216\times$
more particles and $80\times$ more volume compared to our previous
studies in \citet{dav99} and \citet{dav01b}, and they include more
sophisticated physical processes compared to those works and more
recent studies such as \citet{pas09}.  For example, our simulations
now include metal-line cooling and a well-constrained heuristic
model for galactic winds that enrich the IGM and regulate star
formation in accord with observations.  While we will not discuss
chemical enrichment in this work~\citep[see e.g.][]{opp08}, the
energy input from such outflows can non-trivially impact the thermal
conditions in the IGM, particularly around galaxies.  \citet{tor10}
investigated the IGM in simulations similar to ours with the further
addition of feedback from black holes, showing that it can significantly
impact WHIM properties.

In \S\ref{sec:sims} we describe our cosmological hydrodynamic
simulations and the generation and analysis of artificial spectra.
In \S\ref{sec:igmphys} we describe the physical conditions of the IGM
from $z=2\rightarrow 0$, including how \lya\ absorbers trace large-scale
structure and the emergence of WHIM gas.  In \S\ref{sec:stats}
we present statistics of \lya\ absorbers and their evolution.
We also present comparisons to STIS data and predictions for COS,
and study the impact of differences in spectral resolution and noise.
In \S\ref{sec:phys} we discuss how \lya\ absorbers trace the physical
conditions of the low-$z$ IGM, including wide absorbers that trace
WHIM gas.  We summarize and discuss the implications of our results
in \S\ref{sec:summary}.

\section{Simulations} \label{sec:sims}

\subsection{The Code and Input Physics}

We employ our modified version \citep{opp08}
of the N-body+hydrodynamic code \gad \citep{spr05}.
\gad\ uses a tree-particle-mesh algorithm to compute
gravitational forces on a set of particles, and an entropy-conserving
formulation \citep{spr02} of Smoothed Particle Hydrodynamics 
\citep[SPH;][]{gin77,luc77}
to simulate pressure forces and shocks on the gas particles.  We
include radiative cooling from primordial gas assuming ionization
equilibrium~\citep[following][]{kat96} and metal-line cooling based
on \cite{sut93}.
Star formation follows a \citet{sch59} Law calibrated
to the Kennicutt (1998ab) relation; particles above a density threshold
where sub-particle Jeans fragmentation can occur are randomly
selected to spawn a star with half their original gas mass.  The
interstellar medium (ISM) is modeled through an analytic subgrid
recipe following \citet{mck77}, including energy returned from
supernovae~\citep{spr03a}.

Kinetic outflows are also included emanating from all galaxies that are
forming stars.  Gas particles eligible for star formation are randomly
selected to be in an outflow.  Outflowing particle have their velocity
augmented by $v_w$ in a direction given by {\bf v}$\times${\bf a},
where {\bf v} and {\bf a} are the comoving velocity and acceleration of
the particle, respectively.  The ratio of the probabilities to be in an
outflow relative to that for forming into stars is given by $\eta$, the
mass loading factor.  Hydrodynamic forces on wind particles are ``turned
off" until the particle reaches one-tenth the threshold density for star
formation, or a maximum time of 20~kpc/$v_w$.  This hydrodynamic delay
allows wind particles to escape their host galactic disks, mimicking the
effect of ``chimneys'' in the interstellar medium that are unresolvable
in cosmological simulations but would allow the escape of gas in real
galaxies.  The choices for $v_w$ and $\eta$, including their possible
dependence on galaxy or halo properties, define the ``wind model".

We consider three wind models.  The first has no winds, ``nw".  This model
fails a number of observational tests, including overproducing cosmic
star formation~\citep[e.g.][]{dav01,bal01} and failing to enrich the
IGM~\citep{opp06}, but it serves as a baseline model to assess the impact
of outflows on the diffuse IGM.  

The second wind model uses constant values of $\eta=2$ and $\vw=680\kms$
for all outflows from every galaxy; we call this our ``constant
wind" (cw) model.  The cw model is similar to that used in
\citet{spr03b}, \citet{tor10} (though their wind speed is somewhat
lower), and the OverWhelmingly Large Simulations~\citep[OWLS;][]{wie09}.
This model uses all the supernova energy (assuming a Chabrier IMF)
to drive kinetic winds.  Doing so suppresses cosmic star formation
to be broadly in agreement with observations, with the added benefit
of yielding resolution-converged results~\citep{spr03b}.  In detail,
it fails to enrich the IGM in agreement with high-$z$ \ion{C}{iv}
absorbers~\citep{opp06} or to enrich galaxies in accord with
observations of the mass-metallicity relation~\citep{dav07}, but
it is still a physically reasonable scenario that captures some of
the main effects implied by the observations.

Our preferred model, which we refer to as ``vzw", uses scalings expected
for momentum-driven winds~\citep{mur05}, though we note that such
scalings can be generated by other physical mechanisms~\citep{dal08}.
Here, $v_w = 3 \sigma \sqrt{f_L-1}$, and $\eta=\sigma_0/\sigma$, where
$\sigma$ is the velocity dispersion of the host galaxy.
We estimate
$\sigma$ from the galaxy's
stellar mass following \citet{mo98}, using an on-the-fly friends-of-friends
galaxy finder, as described in \citet{opp08}.  The
scaling factor $f_L$ is randomly chosen for each ejection event in the range
$1.05-2$, based on observations of local starbursts by
\citet{rup05}.  This choice also gives outflow velocities consistent
with that seen in $z\sim 1$~\citep{wei09} and $z\sim 2-3$~\citep{ste04}
star-forming galaxies.  The ejection 
speeds are generally comparable to the galaxy
escape velocities at low redshifts, but because wind particles
are slowed by hydrodynamic interactions with surrounding gas,
they typically do not escape their host
galactic halos, at least in more massive
systems.  $\sigma_0$ is a free parameter adjusted to broadly match
the cosmic star formation history; here we choose $\sigma_0=150\kms$.
This yields $\eta\sim 1$ for $L\sim L_*$ galaxies at $z\sim 2$, which
is in agreement with the constraints inferred from $z\sim 2$ galaxies by
\citet{erb08}.


The key features of our momentum-driven wind model are the higher
mass-loading factors and lower ejection velocities in lower mass galaxies.
This model has displayed repeated and often unique
success at matching a range of cosmic star formation and enrichment
data.  This includes observations of IGM enrichment as seen in $z\sim
2-4$ \ion{C}{iv} absorbers~\citep{opp06,opp08}, $z\sim 0$ \ion{O}{vi}
absorbers~\citep{opp09}, and $z\sim 6$ metal-line absorbers~\citep{opp09b}.
Concurrently, it also matches the galaxy mass-metallicity
relation~\citep{dav07,fin08}, and the enrichment levels seen in $z=0$
intragroup gas~\citep{dav08}.  It also suppresses star formation in
agreement with high-redshift luminosity functions~\citep{dav06} and
the cosmic evolution of UV luminosity at $z\sim 4-7$~\citep{bou07}.
In contrast to the nw and cw models, the vzw model approximately
reproduces the observed $z=0$ stellar mass function of galaxies up
to luminosities $L\sim L_*$, though it still predicts excessive
stellar masses for higher mass galaxies \citep{opp10}.
The numerical implementation of this model --- with probabilistic,
one-at-a-time particle ejection and the temporary turn-off of
hydrodynamic forces --- is only a heuristic representation of the
underlying physics, and there are significant numerical and
physical uncertainties associated with the hydrodynamic interactions
between wind and ambient-halo particles and the mixing of outflows
with the surrounding gas.  Nonetheless, this heuristic model
appears to distribute matter and energy into
the IGM and regulate galaxy growth in broad accord with observations.

While we leave a detailed discussion of metals to a forthcoming paper in
this series, the simulations
include a sophisticated chemical evolution model that
follows four individual species (C, O, Si, Fe) from three sources: Type
II SNe, Type Ia SNe, and stellar mass loss from AGB stars.  The former
instantaneously enriches star-forming particles, whose metallicity can
then be carried into the IGM via outflows (described below).  Type~Ia
modeling is based on the fit to data by \citet{sca05}, with a prompt
component tracing the star formation rate and a delayed component (with
a 0.7~Gyr delay) tracking stellar mass.  Stellar mass loss is derived
from \citet{bru03} population synthesis modeling assuming a \citet{cha03}
IMF.  Delayed feedback adds energy and metals to the three nearest gas
particles; Type~Ia's add $10^{51}$~ergs per SN, while AGB stars add
no energy, only metals.  We use \citet{lim05} yields for Type~II SNe;
yields for Type~Ia and AGB stars come from various works~\citep[see][for
details]{opp08}.

\subsection{Simulations}

The primary simulations used here are those described by \citet{opp10},
using the three outflow models outlined above.  Each simulation contains
$384^3$ gas and $384^3$ dark matter particles, in a random periodic
volume of $48\hmpc$ (comoving) on a side with a gravitational softening
length (i.e. spatial resolution) of $2.5\hkpc$ (comoving, Plummer
equivalent).  We assume a cosmology concordant with WMAP-7 constraints
\citep{jar10}, specifically $\Omega_m=0.28$, $\Omega_\Lambda=0.72$,
$h\equiv H_0/100\,\kmsmpc = 0.7$, $\Omega_b=0.046$, $n=0.96$, and
$\sigma_8=0.82$.  This yields gas and dark matter particle masses of
$3.56\times 10^7 M_\odot$ and $1.81\times 10^8 M_\odot$, respectively.
The initial conditions are generated with an \citet{eis99} power spectrum
at $z=129$, in the linear regime, and evolved to $z=0$; they are identical
for all simulations.  Our naming convention is r[boxsize]n[number
of particles per side][wind model], where the initial letter ``r''
indicates the particular choice of cosmology above.  Hence our models
are r48n384nw, r48n384cw, and r48n384vzw15, where the postscript 15 on
vzw indicates our choice of $\sigma_0=150\kms$.

To investigate the effects of the numerical resolution and box size,
we also employ a simulation with $512^3$ gas and $512^3$ dark matter
particles, in a random periodic volume of $96\hmpc$ (comoving) on
a side with a gravitational softening length of $3.75\hkpc$
(r96n512vzw15).  This simulation is run with the same momentum-driven
wind model as our $48\hkpc$ case, and we will refer to it as
``vzw-96".  Compared to vzw-48, it has an eight times larger
simulation volume but 3.375 times worse mass resolution.  Hence it
tests both sensitivity to box size and numerical resolution, although
the dynamic range in either sense is not dramatically large.

\subsection{Generating Spectra}\label{sec:spectra}

Spectra are generated using our spectral generation code {\tt specexbin},
which casts lines of sight at an angle to the simulation axes, allowing
continuous and non-repeating lines of sight to be generated through the
periodic simulation volumes.  We generate 70 such continuous spectra from
$z=2$ to $z=0$.  {\tt specexbin} calculates optical depths in \ion{H}{i}
and a variety of metal ions; here we will only be concerned with \lya.
For line statistics we will typically consider three redshift ranges,
namely $z=0-0.2$ (which we will call ``$z=0$"), $z=0.9-1.1$ (``$z=1$")
and $z=1.8-2$ (``$z=2$").  In each interval, the total path length for
the 70 spectra is $\Delta z=14$.  This is comparable to the path length
to be obtained by, e.g., the program of the COS GTO team.

The procedure for calculating optical depths is
described by \citet{opp08}.  In brief, the algorithm is to (i)
calculate the neutral fraction of each gas particle based on the density,
temperature, and the assumed photoionizing flux (discussed below); (ii)
smooth the neutral hydrogen density 
onto randomly-chosen lines of sight in real
space; (iii) calculate \ion{H}{i} optical depths assuming ionization
equilibrium and optically-thin gas using CLOUDY96~\citep{fer98}; and (iv)
use the \ion{H}{i}-weighted peculiar velocity as well as Hubble flow to
rebin the real space optical depths into redshift space.  This produces
a table of optical depths versus redshift along each line of sight.
Concurrently, for each ionic absorption species {\tt specexbin} 
outputs the optical depth-weighted
temperature, density, and metallicity, which are used to
quantify the physical conditions in each identified absorber.
For lines of sight that parallel box axes, {\tt specexbin} produces
the same \lya\ absorption spectra as 
{\tt tipsy}\footnote{
{\tt http://www-hpcc.astro.washington.edu/tools/tipsy/tipsy.html}}
\citep{her96}.
As reported by \cite{pee10}, the \lya\ flux power spectra computed from
{\tt tipsy} spectra at $z=3$ are indistinguishable from those produced
by the independent spectral extraction code of
\cite{lid09}, confirming the robustness of the procedure to the
details of numerical implementation.

Extracted spectra
are converted from optical depths to
fluxes, resampled to a pixel scale of $6\kms$, and
convolved with the COS line spread function\footnote{
http://www.stsci.edu/hst/cos/performance/spectral\_resolution} (LSF).
Specifically, we use the FUV G130M LSF at 1450\AA.  The LSF varies only
very slightly with wavelength in the FUV channel, but is substantially
different (and narrower) in the NUV channel.  Nevertheless, we keep the
LSF fixed at all wavelengths so as to avoid introducing any artificial
evolution from the varying LSF.  Generally, the FUV LSF is somewhat
broader than had been originally expected, with its central peak having
an equivalent $1\sigma$ Gaussian width of $\approx 17\kms$, and it also has
substantial non-Gaussian wings.  Finally, we add Gaussian random noise
to the simulated spectra assuming a signal to noise ratio of S/N=30
per pixel.  This is comparable to the highest quality data that will be
obtained with COS.

\begin{figure}
\vskip -0.4in
\setlength{\epsfxsize}{0.6\textwidth}
\centerline{\epsfbox{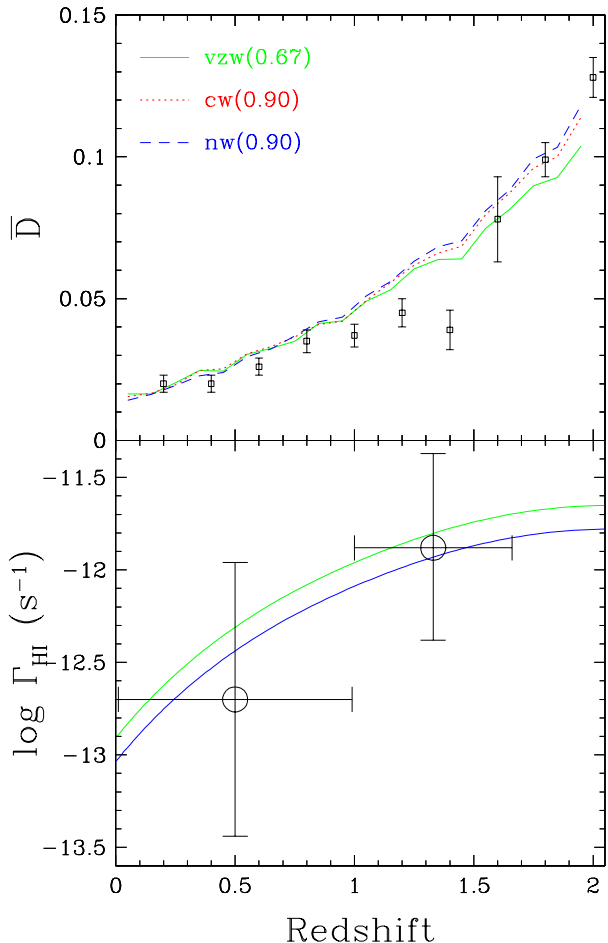}}
\vskip -0.2in
\caption{{\it Top:} Mean flux decrement ($\bar{D}$) evolution from $z=2\rightarrow 0$ in
our three simulations.  Data points are from \citet{kir07}.
The numbers in parantheses are the values of $f_\tau$ by which the
optical depths are multiplied in each simulation to
obtain the agreement with data as shown.
{\it Bottom:} $\ghi$ evolution that matches the
mean flux for the vzw (green) and nw (blue) runs.  The cw run (not shown)
is identical
to nw.  These are simply the \citet{haa01} values divided by $f_\tau$.  
Data points show proximity effect measurements from \citet{sco02}.
}
\label{fig:meanflux}
\end{figure}

The shape and amplitude of the metagalactic photoionizing flux are still
not well constrained, particularly at lower redshifts.  We assume that
the flux is spatially-uniform, which is probably a good approximation in
the diffuse and optically thin \lya\ forest.  We take the spectral shape
from that predicted by \citet{haa01} using the population of observed
quasars and star-forming galaxies
(their ``Q1G01" model).  This model assumes a 10\% escape fraction
of ionizing radiation from galaxies and is broadly consistent with
available constraints~\citep[e.g.][]{sco02}.

\begin{figure*}
\vskip -2.2in
\setlength{\epsfxsize}{0.75\textwidth}
\centerline{\rotatebox{270}{\epsfbox{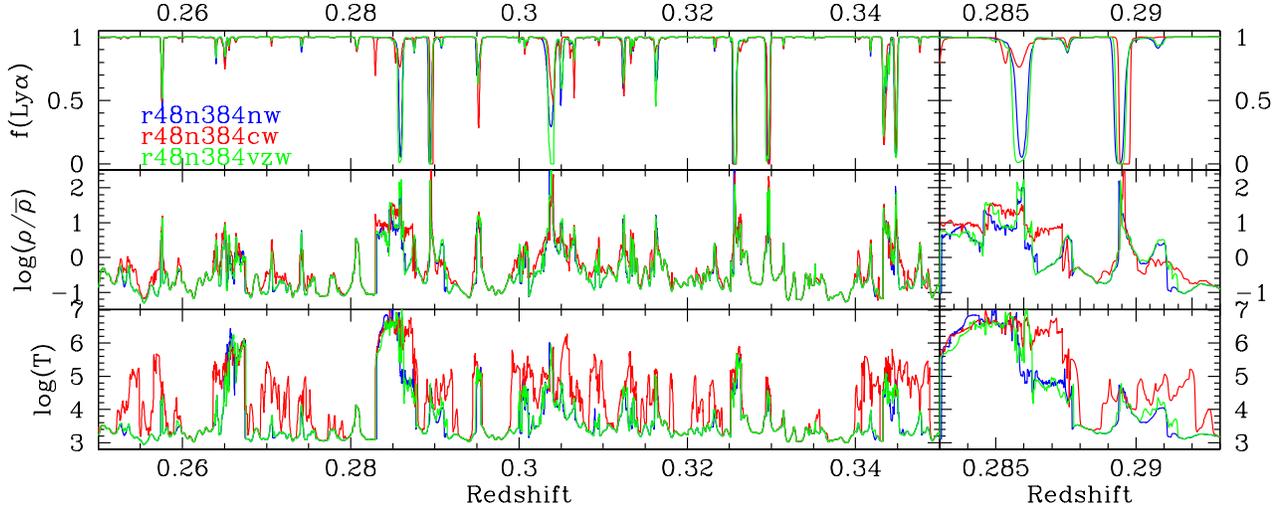}}}
\vskip -0.3in
\caption{Example of a section of noise-free simulated spectra ($\Delta
z=1$), with an expanded region shown on the right, comparing three of our
outflow models.  Top panel shows \lya\ absorption, middle shows density
relative to the cosmic mean, and bottom shows temperature.  The color code is
nw (blue) plotted
on bottom, cw (red) above, and vzw (green) on top; e.g. regions
where only green shows means the others are identical to within the
display resolution.  The density structure is similar in all models,
with the vzw having slightly more absorption overall.  The cw model
clearly produces a warmer IGM and the most deviations in optical depth; 
the other two show smaller differences.
}
\label{fig:spectra}
\end{figure*}

We adjust the amplitude of the metagalactic flux to match the
observed evolution of the mean flux decrement ($\bar{D}$) in the
\lya\ forest.  A modest post-processing adjustment recovers essentially
the identical answer to running the entire simulation with a different
ionizing background, since the dynamics of \lya\ forest gas are
negligibly affected by photoionization~\citep{wei98}.  The amount
of adjustment varies slightly among our simulations, since the
outflows have a small but non-negligible impact on the \lya\ forest.
In practice, we multiply all the optical depths by a factor $f_\tau$
before we convert to flux, then measure $\bar{D}$, and iteratively
adjust $f_\tau$ until we obtain a good match to $\bar{D}(z)$.  In
the optically-thin approximation (excellent in this case), this
procedure is equivalent to multiplying the strength of the ionizing
background by $1/f_\tau$.  While in principle we could have $f_\tau$
be a function of redshift, this was not seen to be necessary to
within the accuracy of available data, and so we simply determine
a constant $f_\tau$ individually for each simulation.

The resulting evolution of the mean flux decrement in the \lya\
forest is shown in the top panel of Figure~\ref{fig:meanflux}.  The
values of $f_\tau$ are indicated in the parentheses within the
legend; they are close to unity, indicating that the original
\citet{haa01} metagalactic flux amplitude is viable to within current
observational uncertainties.  In all cases, the resulting $\bar{D}(z)$
is a reasonable match to observations by \citet{kir07}, who measured
the flux decrement evolution using data compiled from {\it Hubble}'s
Faint Object Spectrograph (FOS) at $z\la 1.5$ and the Keck telescope's
High Resolution Echelle Spectrograph (HIRES) at $z\ga 1.5$.  The
simulations appear to have a somewhat higher flux decrement at $1\la
z\la 1.5$, but in this regime the low resolution of FOS and the
increased blending owing to a thicker \lya\ forest (as compared to
$z\sim 0$) are more likely to result in continuum placement errors
that would cause the flux decrement to be biased low.  Hence we do
not consider these discrepancies serious as yet, though they are
worth revisiting when better data becomes available from COS.

The bottom panel of Figure~\ref{fig:meanflux} shows the evolution of
the \ion{H}{i} photoionization rate $\ghi$ that yields the mean flux
evolution in the top panel.  Since we multiply the \citet{haa01} model by
a constant factor, the shapes of the curves are identical to their model.
The amplitude is multiplied by $f_\tau^{-1}$, so vzw is slightly higher
than nw; cw is identical to nw since it has the same $f_\tau$.  All the models
are consistent with the available observational constraints as they are
fairly uncertain \citep[see][for more constraints]{dav01b}; here we show
proximity effect measurements from \citet{sco02}.  If we take the $z=1.4$
measurement of $\bar{D}$ at face value and modify $\ghi(z)$ to match it,
this would result in an increase in $\ghi$ by 0.2~dex at $z=1.4$, and a
subsequent recovery back to the original track by $z=1.6$.  This would
be an odd detour in its evolution, implying a peak in photoionizing flux
at $z\sim 1.5$ along with a secondary peak at $z\sim 2-2.5$ as in
the original \citet{haa01} evolution.  In this case the peak epoch of
$J_\nu$ would no longer correspond to the peak epoch of high-luminosity
quasars, although lower luminosity quasars peaking at lower redshifts
could provide a significant contribution~\citep{sha09}.

It is interesting that the constant wind and no wind cases have
identical values of $f_\tau$, while the vzw case is lower.  In the
simplest view, winds tend to add energy to the IGM, resulting in
less \lya\ absorption, and hence one needs smaller optical depths
and a lower $f_\tau$.  In detail however, because $\bar{D}$ is
dominated by marginally saturated absorbers, the relevant issue is
whether the winds are heating gas giving rise to those absorbers
in particular.  It turns out that vzw, owing to its smaller wind
speeds from typical galaxies, tends to deposit energy closer to
galaxies where strong absorbers arise, while cw tends to deposit
energy far away in more diffuse gas, as we will discuss in
\S\ref{sec:igmphys}.  Hence we believe that although cw adds more
energy to the IGM overall, this does not impact $\bar{D}$ nearly as
much as in vzw.

Figure~\ref{fig:spectra} shows an example spectrum from $z=0.3-0.4$
along an identical line of sight (LOS) through our three wind
simulations.  The \lya\ forest is fairly sparse at these epochs,
but many very weak lines exist that could be detected given sufficient
observational capabilities.  The mean absorption is similar by
construction, but there are distinct variations evident particularly
between the cw case and the other two.  The expanded segment in the
right panels highlights a region that shows significant differences
among all three wind models.  The density structure is mostly
independent of outflows, though small differences are evident, but
the IGM temperature can be significantly impacted by wind energy.
The constant wind case shows a higher temperature almost everywhere
along the LOS, while the temperatures for vzw and nw are fairly
similar.  However, the fact that optical depths are not very sensitive
to temperature makes these differences difficult to pick out in the
observed spectra.  Overall, winds do not have a strong impact on
the physics or absorption properties along typical lines of sight,
as we will demonstrate more quantitatively with \lya\ absorber
statistics.

\section{IGM Physical Conditions}\label{sec:igmphys}

\subsection{Cosmic Phase Diagram}

\begin{figure*}
\vskip -0.6in
\setlength{\epsfxsize}{0.9\textwidth}
\centerline{\epsfbox{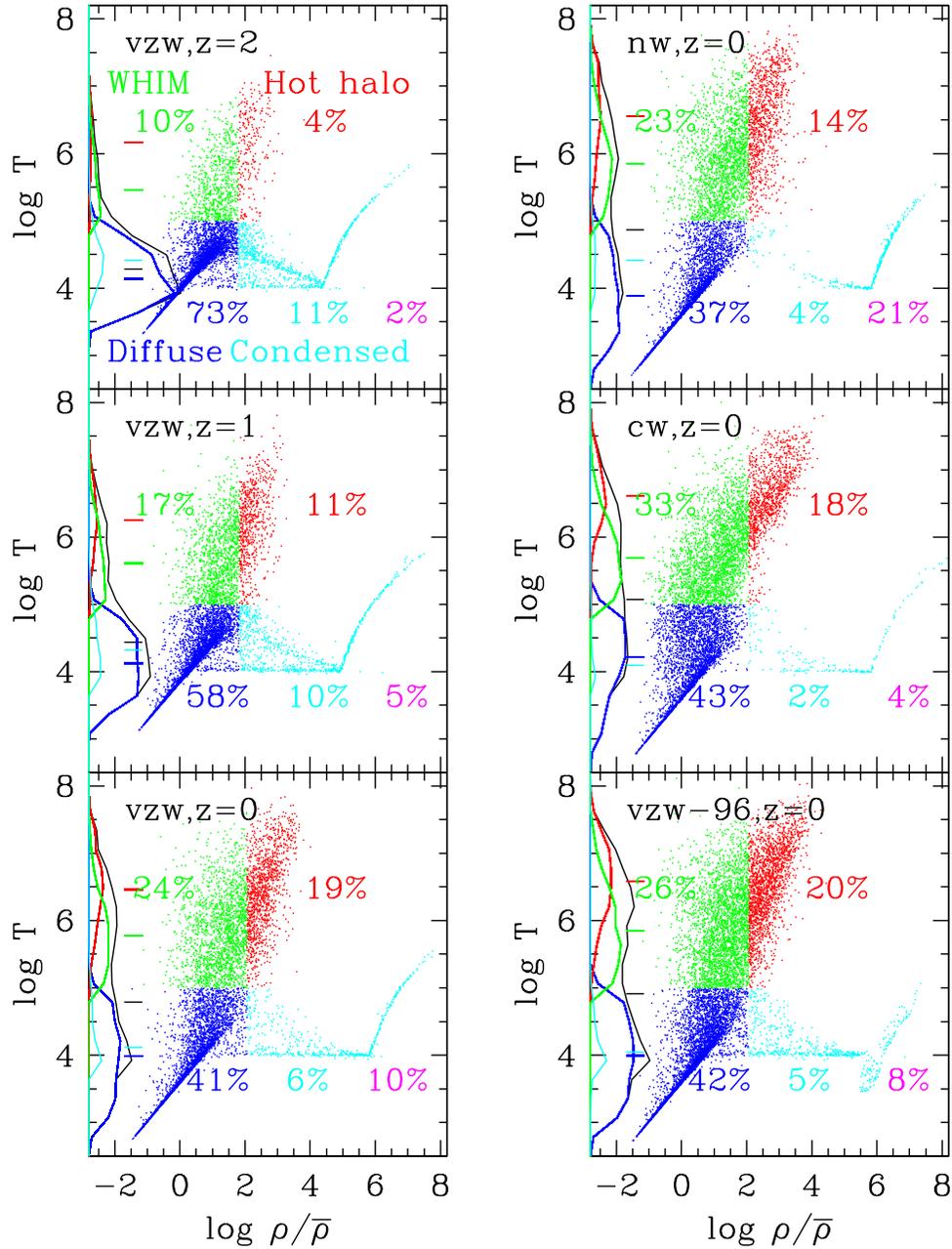}}
\vskip -0.7in
\caption{{\it Left panels:} 
Cosmic phase space diagram for gas particles in the vzw run 
at $z=2,1,0$ (2000 randomly-subsampled gas particles).
We divide the baryons into four categories: 
Diffuse, Condensed, Hot Halo, and WHIM.  This is done using a temperature
cut at $10^{5}$~K to separate the former two from the latter two,
and a density cut given by equation~\ref{eqn:deltath} to separate
the bound baryons (condensed and hot halo gas) from the intergalactic
baryons (diffuse and WHIM).  Temperature
histograms are along the $y$-axis for each phase and the total (black),
with median values indicated by the tick marks.  The percentage of mass
in each phase is listed in each panel; additionally, the fraction in 
stars is shown in magenta.
{\it Right panels:} Same, for the other simulations (nw, cw, and vzw-96),
all at $z=0$.
}
\label{fig:rhoT}
\end{figure*}

The left panels of Figure~\ref{fig:rhoT} shows cosmic phase diagrams of
baryons, i.e. overdensity versus temperature, in our fiducial vzw run
at redshifts from $2\rightarrow 0$.  
The trends shown are familiar.  At
the lowest densities there is a tight density-temperature relation set
by a competition between photoionization heating and adiabatic cooling
owing to Hubble expansion.  At roughly the cosmic mean density, shock
heating begins on collapsing filamentary structures, creating a plume of
hotter gas.  At higher densities, radiative cooling becomes effective,
creating the spur of dense gas that fuels star formation in galaxies.
The other wind models look qualitatively similar.

Two perhaps unfamiliar features are the shelf of gas sitting at $T\approx
10^4$K, which is metal-enriched gas outside of star formation regions
(we truncate all cooling at $10^4$K), and the upwards-curving plume of
the densest gas, which is a consequence of the two-phase interstellar
medium implementation of \citet{spr03a}.  The temperature in the
two-phase regime is given by a mass-weighted average of the subgrid hot
phase, assumed to be at $10^8$~K, and the mass-dominant cold phase at
$1000$~K.  Neither of these features are physical; they simply represent
computational conveniences.

From redshift 2 to 0, we see the growth of galaxies and large-scale
structure playing out in the cosmic phase diagram.  The denser phases
become more heavily populated, as does the plume of shock-heated WHIM
gas, though the hotter phases are still clearly populated even at $z=2$.
The hottest gas also grows most substantially owing to the formation of
large potential wells at late times.  The diffuse gas extends into lower
overdensity regions at late times, as voids become more empty, and its
temperature drops with time as the physical density and photoionization
rates go down.  The overall cosmic gas temperature distribution becomes
broader with time, and shifts to slightly higher temperatures, driven by
the increasing fraction in WHIM and hot halo gas.  These trends have all
been noted and described in previous works~\citep[e.g.][]{dav99}.

\subsection{Redefining Cosmic Baryon Phases}
\label{sec:phases}

In Figure~\ref{fig:rhoT} we have also subdivided our baryons into four
cosmic phases.  The subdivisions are slighly different than conventional
ones~\citep[e.g.][]{dav01}, in which diffuse baryons are at $T<10^5$~K,
WHIM gas is $10^5<T<10^7$~K, hot gas is $T>10^7$~K, and condensed baryons
are at a fairly high overdensity.  Our new divisions are intended to
be more physically-motivated, particularly in the case of the WHIM,
though in practice they do not represent dramatic changes.

The main difference is that we divide the phases according to an
overdensity threshold.  This threshold is intended to reflect the
overdensity ($\delta\equiv\rho/\bar\rho-1$) at the boundary of a
virialized halo.  To determine this, we note that for reasonable
choices of the \citet{nav96} concentration parameter $c$, the
density at the virial radius is roughly one-third the mean enclosed
density\footnote{In detail, $\rho(r_{200})/\bar\rho = \frac{1}{3}
\frac{c^2}{(1+c)^2(ln(1+c)-c/(1+c))}$, which is approximately $1/3$ for
low values of concentrations corresponding to objects just undergoing
collapse.}.  We thus obtain~\citep{kit96}
\begin{equation}\label{eqn:deltath}
\delta_{\rm th} = 6 \pi^2 (1+0.4093(1/f_\Omega-1)^{0.9052})-1, \\
\end{equation}
where
\begin{equation}
f_\Omega = \frac{\Omega_m (1+z)^3}{\Omega_m (1+z)^3+(1-\Omega_m-\Omega_\Lambda)(1+z)^2+\Omega_\Lambda}.
\end{equation}
$\delta_{\rm th}$ serves as our division between ``bound" phases (gas in
galaxies and halos) and ``intergalactic" phases (WHIM and diffuse gas).
The value of $1+\delta_{\rm th}$ evolves from $6\pi^2 \approx 60$ at 
$1+z \gg (\Omega_\Lambda/\Omega_m)^{1/3}$ to $\approx 120$ at $z=0$.
This boundary is not completely sharp, as some gas that has higher
overdensities is not yet bound, and owing to the elliptical nature of
halos some bound gas can have lower overdensities than $\delta_{\rm th}$.
We checked this definition by running a spherical overdensity halo
finder~\citep{ker05}, and checking that equation~(\ref{eqn:deltath})
faithfully and robustly separated gas within halos from gas outside.  
It indeed did so, as unbound gas was never more than a couple tenths of 
a dex in density into the bound region, and vice versa.

We further separate the ``hot" phases (WHIM and hot halo gas) from
``cool phases" (diffuse and condensed gas) by a temperature threshold of
$T_{\rm th}=10^5$~K, the same value used in earlier definitions.
At high densities, all gas that is star-forming
is included in the condensed phase, although the subgrid two-phase
model creates a mean gas particle temperature that can exceed $T>T_{\rm
th}$.  At IGM densities, this represents the temperature above which
the \ion{H}{i} neutral fraction starts to drop dramatically, making
\lya\ absorption a poor tracer of its baryonic content (though as we will
see later, wide \lya\ absorbers can still trace this gas).  Furthermore,
since photoionization cannot easily heat gas to these temperatures,
it demarcates gas that has been shock-heated by large-scale structure.
From Figure~\ref{fig:rhoT}, one can see that some gas at 
$T=10^4-10^5$~K is also shock-heated above the photoionization locus,
but the locus itself extends to $T \approx 10^5$~K, at least at $z\geq 1$.

With these divisions, we have four phases as follows: 
\begin{itemize}
\item Diffuse ($T<T_{\rm th}, \delta<\delta_{\rm th}$),
\item WHIM ($T>T_{\rm th}, \delta<\delta_{\rm th}$),
\item Hot halo ($T>T_{\rm th}, \delta>\delta_{\rm th}$),
\item Condensed ($T<T_{\rm th}, \delta>\delta_{\rm th}$),
\end{itemize}
\noindent with the qualification that two-phase star-forming gas 
is ``condensed'' regardless of temperature.
The most significant departure from previous definitions is that
of the WHIM gas.  
Unlike the common definition in which $10^5-10^7$~K gas is
WHIM regardless of density,
the WHIM gas in our definition is truly
intergalactic, in the sense that it is not within the boundaries
of a galaxy or its halo.
Also, since hot gas is most commonly detected in X-ray
emission, which roughly scales as $\propto \rho^2T^{1/2}$ (neglecting
metal line emission for the sake of this argument), this lower-density
gas is more likely to be ``missing" from current observational
censuses.  For instance, our definition removes hot gas in galaxy groups at
$T\la 1$~keV from being part of the WHIM; such gas can be detected
in soft X-rays~\citep{mul00} because its high density 
(relative to truly intergalactic gas)
leads to significant emission, albeit at faint levels.
Hence, our revised definition more closely reflects the idea that
the WHIM is the repository of the ``missing'' cosmic baryons.
Eliminating the upper limit of $T=10^7$~K for the WHIM has minimal
impact, since very little gas with $\delta < \delta_{\rm th}$ has
such high temperatures.

\subsection{Winds and the Phase Diagram}


The right panels of Figure~\ref{fig:rhoT} show the $z=0$ cosmic phase
diagram for our other simulations: nw, cw, and vzw-96.  Comparing vzw-96 to vzw,
we see some subtle differences particularly in the high-temperature gas.
The hot halo gas extends to slightly higher temperatures, owing to
larger halos being able to form in the $96\hmpc$ volume; this causes a
slightly higher median temperature for this phase, from $10^{6.46}$~K
to $10^{6.55}$~K.  Similarly, the WHIM gas extends to slightly higher
temperatures, causing a comparable increase in median WHIM temperature.
The cooler phases show no appreciable physical differences.  There is
also a feature in the condensed phase at $\rho/\bar\rho\approx 10^6$
corresponding to a slightly different implementation of the two-phase
cutoff in this run, but this does not affect any of our results.

The no-wind case also shows some differences with respect to vzw.
Although the distribution of WHIM gas in phase space is similar, the
median temperature increases slightly from $10^{5.77}$~K to $10^{5.85}$~K, 
just as in the vzw-96 case but for a different reason.  Here, the
main effect is that with no outflows, the IGM is almost completely
unenriched~\citep{opp08}.  Hence there is no metal cooling in this
regime, which yields somewhat higher temperatures.  For the same reason,
the condensed phase shows almost no particles at the $10^4$~K metal
cooling floor.

The constant wind model noticeably pushes WHIM gas into a lower-density
regime compared to vzw, but its median temperature is lower
($10^{5.69}$~K) since the diffuse IGM is more enriched~\citep{opp06}.
In contrast, the hot phase temperature is higher by $\approx
0.15$~dex, reflecting the significant IGM heating that occurs when
winds from all galaxies, even small ones, are emanating at $680\kms$.
These results are qualitatively similar to those from \citet{tor10}
who explore a similar wind model, and they further note that black
hole feedback (as they implement it) causes even more heating of
particularly low-density gas.  It is also similar to results from
mesh hydrodynamic simulations by \citet{cen06}, who found that winds
increase the mass fraction in the WHIM by 20\%.  Hence, it is
possible to non-trivially impact the cosmic distribution of baryons
in phase space with plausible outflows.

\subsection{Baryon Phase Evolution}

\begin{figure}
\vskip -0.5in
\setlength{\epsfxsize}{0.6\textwidth}
\centerline{\epsfbox{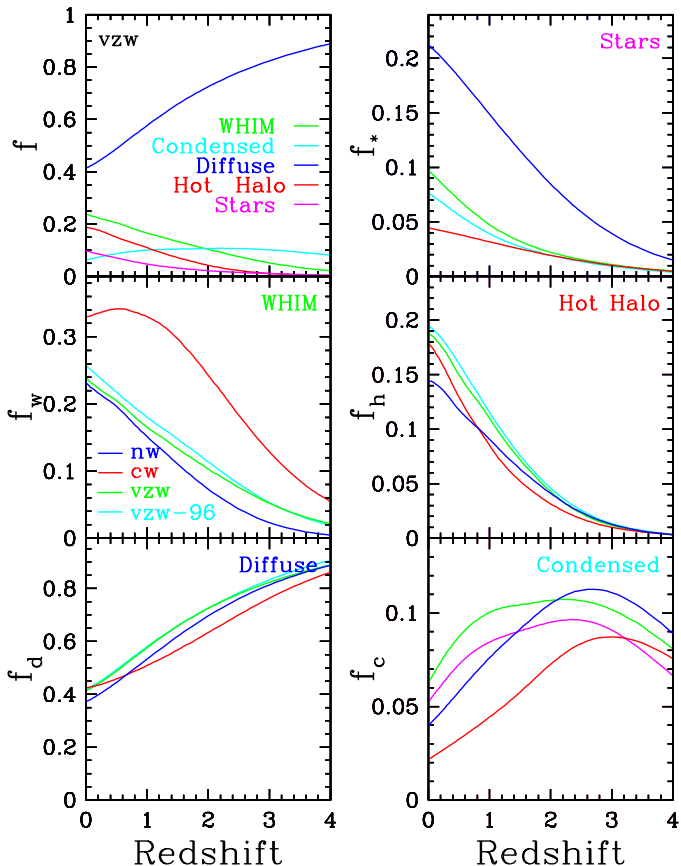}}
\vskip -0.5in
\caption{{\it Top left:} Evolution from $z=4\rightarrow 0$ of baryon
fraction in various cosmic phases, including stars, in our fiducial
vzw run.  {\it Next 5 panels:} Evolution of baryon fraction in stars
(upper right), and then the various baryonic phases in order of their
location in $\rho-T$ space: WHIM (middle left), hot halo (middle right),
diffuse (lower left), and condensed (lower right), in our three wind
models (green, blue, red lines), along with the larger-volume vzw-96 simulation
(cyan).
}
\label{fig:baryphase}
\end{figure}

We now quantify the trends in the cosmic phase diagram by examining
the evolution of baryons in each phase amongst our various simulations.
As is apparent from Figure~\ref{fig:rhoT}, baryons progressively
move from the diffuse phase into WHIM and hot halo gas, owing to the
hierarchical growth of large potential wells.  The condensed phase gas
is supplied predominantly from cold mode accretion but also some cooling
from halo gas~\citep{ker09}, and this provides the fuel to grow the
cosmic stellar content.  By $z=0$, baryons are comparably divided
between diffuse, WHIM, and the various bound phases, as was found in
\citet{dav01}.  However, the new definition of the WHIM yields somewhat
less WHIM gas and more diffuse gas, tilting the balance towards diffuse
gas overall.  The precise balance between the diffuse and WHIM phases
is also sensitive to the adopted value of $T_{\rm th}$, as one can see
from Figure~\ref{fig:rhoT}.

Figure~\ref{fig:baryphase}, top left panel, shows the fraction of
baryons in each cosmic phase from $z=4\rightarrow 0$ in our vzw run.
We additionally show the evolution of baryons in stars.  Qualitatively,
the behavior is very similar to that seen in all cosmological hydrodynamic
simulations~\citep{dav01}, as discussed above.  The next five panels
of Figure~\ref{fig:baryphase} show a comparison of individual phase
evolution for our three wind prescriptions, plus our large-volume
vzw-96 run.  The upper right panel shows the evolution of the global
stellar content, i.e. the integral of cosmic average star
formation rate density vs. redshift~\citep{mad98}, in each wind run.
Note that these values account for stellar evolution which returns
material back into the gas phase.  The no-wind case produces by far the
most stars, which is the well-known result that without strong feedback,
there is a severe overcooling problem~\citep{dav01}.  A stellar baryon
fraction today of $\sim 20\%$ far exceeds current observational values of
$<10\%$~\citep[e.g.][]{bel03}.  Though we don't show it here, this case
is also poorly resolution-converged, as pointed out by \citet{bal01},
and hence the problem would be worse at higher resolution.

By contrast,
both wind models do a good job of suppressing star formation.  
Their globally averaged star formation rates are 
almost identical down to $z\sim 2$, which is by design as the
wind prescriptions are tuned to reproduce the observed star formation
rates up to this epoch.
At lower redshifts, there start to be greater differences.  Because 
the vzw model has more recycling of wind material back into galaxies,
it provides more fuel for
late-time star formation~\citep{opp10}.  None of our simulations
includes a prescription (e.g., AGN feedback)
to truncate star formation in the most massive
systems as observed, which would lower the predicted stellar fractions
particularly at late times.  The resolution dependence of
global star formation is mitigated by the presence of outflows, as shown
by \citet{spr03b} for the cw case, and as shown here by comparing vzw and
vzw-96.  Both wind models are generally consistent with the currently favoured
values of the stellar baryon content today of $4-8\%$~\citep{bal08}.

The middle left panel shows WHIM evolution, which displays significant
variations, particularly between models with and without galactic
winds.  The modest wind speeds of the vzw model do not heat much
gas to WHIM temperatures compared to the purely gravitational (nw)
case, but there is still some heating evident at high-$z$, which
diminishes to $z=0$.  The cw case shows a significant amount of
WHIM gas even at $z=4$, and the WHIM fraction increases faster than
in the other cases down to $z\sim 1$.  At this point the supply of
wind energy abates, and the WHIM fraction levels off and actually
falls slightly to $z=0$.  \citet{tor10} found less dramatic effect
on the WHIM fraction with similar simulations, although the energy
input from their winds is half of ours.  These differences highlight
an important point: galactic winds can in principle have a strong
effect on the evolution of WHIM gas.  The winds in our favoured
model do not substantially affect the WHIM, but faster winds in
plausible variant models can do so.  Measuring the evolution of the
global WHIM content could provide constraints on the large-scale
thermal impact of outflows. Such measurements will be challenging
given the difficulty of quantifying the WHIM even at the present
epoch, but the redshifting of some UV tracer lines to longer
wavelengths could potentially facilitate such measurements.

The middle right panel shows hot bound gas evolution.  In general this
is dominated by gravitational heating from large potential wells, as
indicated by the no-wind model.  The vzw case produces more hot bound
gas at late times, as it removes material from galaxies and places it
into their surrounding hot halos.  Since the wind speed scales with
escape velocity, large galaxies can have outflow velocities that exceed
the wind speed of the cw model, which provides greater heating in their
large halos.  The cw model removes a significant amount of hot gas at
early times because its high-speed winds can escape smaller potential
wells, but at late times the potential wells become deep enough to retain
the gas, and its hot phase rises more quickly.  Still, even by $z=0$
this model does not produce as much hot bound gas as the vzw model.
The vzw-96 and vzw simulations are very similar again, showing that
simulation volume effects are not important even for large potential
wells in terms of their global mass content.

In the lower left panel, the diffuse phase shows similar trends in
all the models.  The vzw winds have little impact on the diffuse gas
content, as their typically modest velocities do not produce substantial
IGM heating.  The cw winds lower the diffuse gas fraction at high-$z$
by shock heating some IGM gas above $10^{5}$~K.  This heating produces
noticeable broadening of high-$z$ metal lines, leading to worse agreement
(relative to vzw) with observed \ion{C}{iv} line widths \citep{opp06}.
At late times, heating of the low density IGM declines because of the
reduced global star formation and the trapping of the constant-$v_w$
winds by deeper potential wells, causing the diffuse phase fraction to
return to the no-wind value by $z\sim 0.5$.  The box size has a negligibly
small impact on this phase --- the cyan vzw-96 curve is hidden by the
overlaid green vzw curve at all but the highest redshifts.

The bottom right panel shows the condensed phase evolution.
The overall trend for this phase is to increase at early epochs as
material accumulates in halos, then decrease towards later times as
galaxies convert gas to stars more rapidly than they accrete fresh gas.
The fraction of condensed gas is $\la 10$\% at all times and $\la 6\%$
today, in all the models.  The cw model shows a fixed offset to lower
condensed gas fractions relative to the no-wind model, since it expels
a constant amount of material out of the star-forming regions, and most
of this material ends up outside of halos owing to the large wind speed.
The vzw model shows qualitatively different behavior from the others,
with less redshift evolution.  Mostly this difference reflects the
late-time reaccretion of wind material, more prominent in the vzw model
because of lower wind speeds \citep{opp10}.  We note the qualitative
similarity of the evolution of this phase (which contains most of the
cosmic neutral gas) with the observed evolution of cosmic \ion{H}{i}
gas from Damped \lya\ systems~\citep{wol05}, being roughly constant from
$z\sim 3\rightarrow 0.5$ and then dropping to $z=0$.

Returning to Figure~\ref{fig:rhoT}, the histograms along the vertical axis
show the temperature distribution of baryons at $z=2,1,0$.  The overall
trend follows the movement of baryons within the cosmic phase diagram,
namely that the median temperature increases with time.  In detail,
the diffuse phase actually moves towards lower temperatures
because of the changing balance between photionization heating and 
adiabatic cooling.
The WHIM gas extends
up to higher temperatures at low-$z$, and the hot bound gas also becomes
hotter on average as larger potential wells form.  The median cosmic
gas temperature today is $\approx 10^5$~K.

Overall, the cosmic evolution of baryon phases reflects the hydrodynamic
and radiative processes associated with the growth of structure.
Winds have a significant impact, particularly on the stellar phase (by
design) and on the WHIM phase (if the winds are strong enough).  They also
impact the bound phases 
(hot halo and condensed), since these lie close to galaxies where winds
play a more important dynamical role.  The diffuse phase is mostly
unaffected by winds.  The complex dynamical interplay between outflows
and hierarchical growth is an important factor in the cosmic history
of baryons, and improved observational constraints and theoretical
predictions will be required to fully characterise it.

\section{\lya\ Absorber Observables}\label{sec:stats}

We now investigate the statistics of \lya\ absorbers in our simulated
COS-like spectra, and present comparisons when appropriate to
existing pre-COS data.  We consider the canonical statistics of the
column density and linewidth distributions, as well as the evolution
of the number density of lines.  To obtain line statistics, we fit
our simulated spectra with Voigt profiles using \autovp~\citep{dav97},
yielding an \ion{H}{i} column density ($\nhi$), Doppler line width
($b$), and redshift for each absorber.  Owing to the sparseness of
the low-$z$ \lya\ forest, the fits are generally unambiguous, and
tests have shown that \autovp\ obtains line parameters very similar
to that from other Voigt profile fitters (e.g. VPFIT) when the lines
are unsaturated.  We use the continuum directly provided by the
simulations, as we assume that observed spectra can be fit with an
accurate continuum in the relatively sparse low-$z$ \lya\ forest.
We also examine the impact of varying the assumed signal-to-noise
ratio and spectral resolution of our simulated spectra.  Note that
AutoVP provides ``raw'' $b$-parameters uncorrected for instrumental
resolution.  In our COS-resolution spectra, with a Gaussian width
of $17\kms$, we therefore expect essentially no lines with $b <
17\sqrt{2} \approx 24\kms$.  Since it turns out that the vzw-96
absorber statistics are indistinguishable from the vzw case (as
anticipated by the close agreement in IGM physical properties), we
do not show this model herein.

\subsection{Column Density Distributions}\label{sec:cdd}

\begin{figure*}
\vskip -0.8in
\setlength{\epsfxsize}{0.9\textwidth}
\centerline{\epsfbox{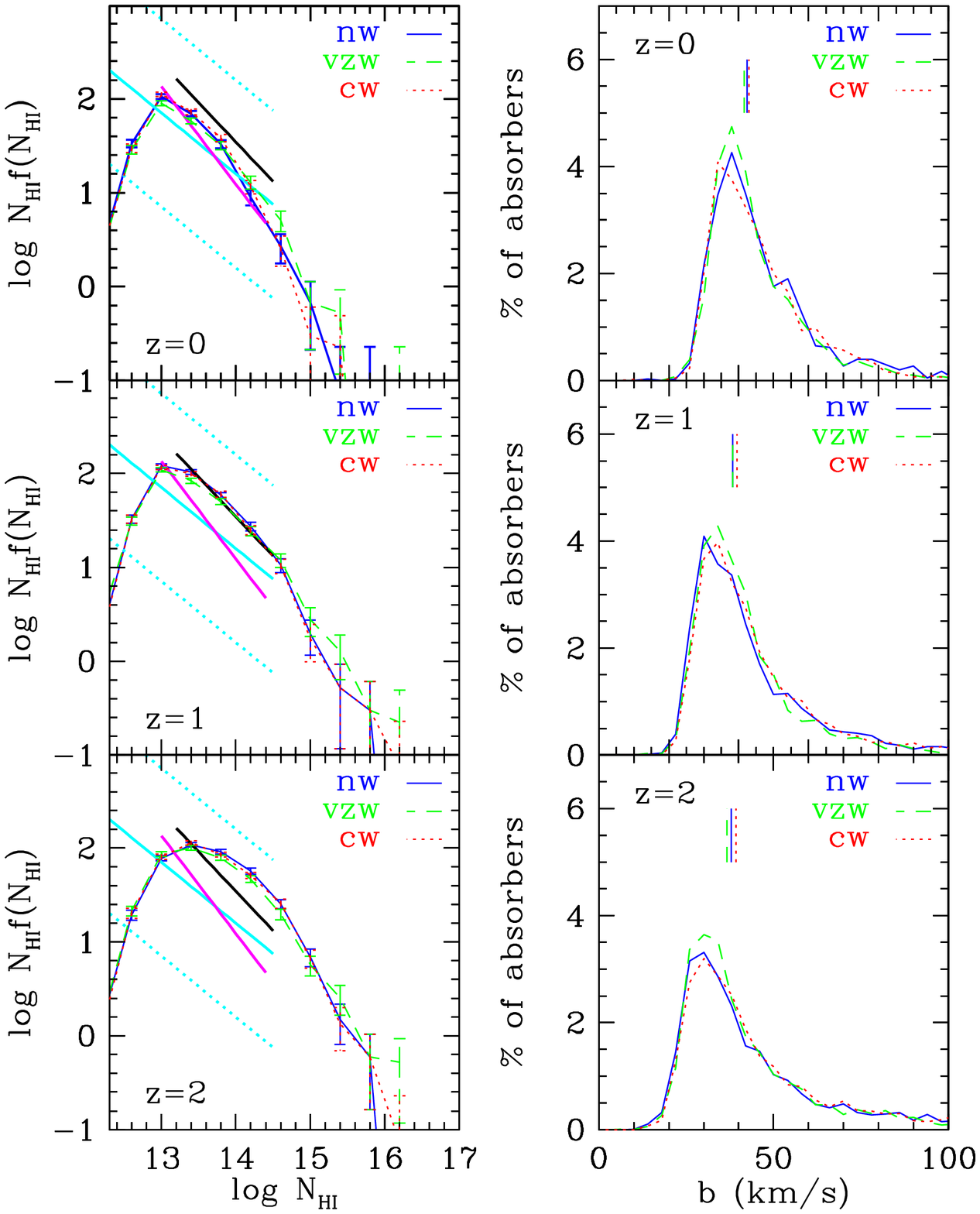}}
\vskip -0.8in
\caption{Column density (left) and $b$-parameter (right) distributions at
$z=0$ (top), $z=1$ (middle), and $z=2$ (bottom) for our three simulations.
The CDD have been multiplied by $\nhi$ improve visibility of the
differences between models.
The thick cyan line shows a fit to data from \citet{pen04} at $z\approx
0$, namely $\log f(\nhi)= 10.3 - 1.65\log{\nhi}$,
over their fitted range of $12.3\leq \log{\nhi}\leq 14.5$;
dotted lines on either side indicate their $1\sigma$ quoted uncertainty
of $\pm1.0$ in $\log f(\nhi)$.  Thick black and magenta lines show
best fits for CDDs from STIS data from \citet{leh07} and \citet{dav01},
respectively.
The same lines are reproduced at the other redshifts for reference,
showing that there is modest evolution in the CDD amplitude from
$z=2\rightarrow 0$.  The turnover at $\nhi<10^{13}\cdunits$ arises from
incompleteness in recovering lines at the assumed S/N and resolution;
see \S\ref{sec:res}.  Linewidth distributions are shown for all lines with
$\nhi>10^{13}\cdunits$, i.e. our complete sample at COS resolution and an
assumed S/N=30.  Tick marks above each curve indicate median $b$ values.
}
\label{fig:Nb}
\end{figure*}

Figure~\ref{fig:Nb} shows column density (left panels) and $b$-parameter
(right panels) distributions for absorbers at $z=0,1,2$.  As mentioned
previously, these three bins correspond to all lines within redshift
ranges $z=0-0.2$, $z=0.9-1.1$, and $z=1.8-2$, respectively, identified
with \autovp\ in our 70 random lines of sight.  Results are shown for
all three of our outflow models.  The column density distribution (CDD)
$f(\nhi)\equiv d^2N/d\nhi dz$ has been multiplied by $\nhi$ to remove
some of the steep power-law scaling (discussed below), which enhances
the visibility of the differences among models.  It also results in a
dimensionless number that reflects the relative number of lines per unit
redshift at a given $\nhi$.

The CDD shows the usual power-law behaviour, above the completeness
limit at $\nhi\approx 10^{13}\cdunits$ that results from our adopted
S/N and resolution (see \S\ref{sec:res}).  Characterizing the column
density distribution as a power law with $f(\nhi)\propto \nhi^{-\beta}$,
and fitting between $10^{13}<\nhi<10^{14.5}\cdunits$, we obtain
values of $\beta=-1.88, -1.70, -1.80$ for the nw, vzw, and cw models
at $z=0$, respectively, with an uncertainty of approximately $\pm
0.1$ on each.  Hence, all the slopes are formally consistent with
each other, although given that the exact same LOS are being analyzed,
one can conclude that the nw model produces a steeper slope than
the wind models, and that the vzw model produces the shallowest
slope.  We note that all the slopes are shallower than the slope
$-2.04\pm 0.23$ obtained in the simulation by \citet{dav01} (though
within formal uncertainties), which was most analogous to the no-wind
case here.  We will show in \S\ref{sec:res} that higher spectral
resolution tends to produce a somewhat steeper slope, so this could
explain part of the difference.  \citet{pas09} found a slope of
$-1.84\pm0.02$ using a fixed mesh simulation with 75~kpc resolution
and no galactic outflows, which basically agrees with our no-wind
case.

These are several pre-COS set of low-$z$ \lya\ forest data for
comparison.  One sample, from \citet{pen04}, contains 187 somewhat
heterogeneous \lya\ absorbers, mostly at $z\la 0.1$, from {\it
Hubble}/STIS and Goddard High Resolution Spectrograph data.  A more
uniform and higher-resolution sample of 341 absorbers out to $z\la
0.4$ was obtained by \citet{leh07} using only STIS data, and included
{\it Far Ultraviolet Space Explorer} data to help constrain the
properties of saturated \lya\ absorbers.  We also show earlier
results from \citet{dav01}, who analyzed two STIS spectra using
\autovp.  The fits to all their resulting CDDs are shown as the
thick lines in the upper left panel of Figure~\ref{fig:Nb}; black
for \citet{leh07}, magenta for \citet{dav01}, and cyan with
$\pm1\sigma$ uncertainties shown as the dashed lines for \citet{pen04}.
The slope obtained by \citet{pen04} is $-1.65\pm 0.07$, which is
statistically different from that of \citet{leh07} ($-1.84\pm 0.06$)
or \citet{dav01} ($-2.04\pm 0.23$).  As we will show in \S\ref{sec:res},
at least part of the difference likely owes to spectral resolution,
since higher resolution data yields steeper slopes.  For this reason
it is not straightforward to compare to the COS predictions presented
here to these data, but broadly the slope and amplitude predicted
by the current simulations are in agreement given the various
uncertainties.  A careful comparison against upcoming COS data should
provide more discrimination between models.

At higher column densities, \citet{pen04} found that their data
showed a characteristic ``dip" in the CDD at $\nhi\sim 10^{15}\cdunits$,
which they point out persists from high-$z$ (see their Figure~9).
\citet{leh07} likewise saw that the CDD slope is shallower at
$\nhi>10^{14.4}\cdunits$ than for smaller systems.  We see a hint
of the onset of such a dip in all our simulations, but we cannot
accurately trace it out to $\nhi\ga 10^{16}\cdunits$, since we
quickly run out of absorbers owing to the rarity of high-$\nhi$
absorbers along our randomly-chosen LOS.  The origin of this dip
is unclear.  It occurs at a column density close to that expected
for overdensities near the boundaries of galaxy halos, which leads
us to speculate that perhaps these absorbers probe the outer parts
of galaxy halos where (at least in larger halos) some of the gas
might be shock-heated to a temperature where it would not absorb
in \ion{H}{i}.  At even higher columns, one would then probe through
denser halo regions where \ion{H}{i} condenses out again, owing to
self-shielding, causing a ``recovery" back to the original CDD.  We
leave a detailed examination of this conjecture for future work.

The redshift evolution of the CDD shows a steady march upwards in
line counts at a given $\nhi$ ($\ga 10^{13.5}\cdunits$).  The overall
steady evolution reflects the evolution of the mean flux decrement
over this redshift range, and hence is dependent on our procedure
of adjusting the ionizing background strength to match $\bar{D}(z)$.
The incompleteness at $\nhi\la 10^{13.5}\cdunits$
becomes more severe at higher-$z$, as line blending becomes more common.
Recall that we have fixed our resolution and line spread function to the
COS far-UV channel at all $z$, which underestimates the data quality that
is obtainable at $z\sim 1$ (where the COS NUV channel has a narrower LSF,
albeit with lower sensitivity) and at $z\sim 2$ (where one can trace
the \lya\ forest in the optical).  This makes it difficult to compare
slopes across a fixed $\nhi$ range, but if we restrict ourselves to
$10^{13.4}-10^{14.6}\cdunits$ where the CDD is reasonably complete at
all $z$, we obtain slopes for the vzw model of $-1.89, -1.71$, and $-1.54$
at $z=0,1,2$, respectively.  The CDD is clearly getting
steeper with time, broadly matching the observed CDD slope evolution
from high-$z$~\citep[$\sim -1.5$, e.g.][]{kim01,jan06} to low-$z$.

The CDDs are remarkably insensitive to the outflow model at lower
column densities.  Wind dynamics, despite enriching the IGM and
depositing energy over large scales in quite different ways, have
little impact on the diffuse \ion{H}{i} distribution.  As shown by
\citet{opp08} and \citet{kol03,kol06}, the typical extent of galactic
winds ($\sim 100$~kpc physical) generally does not extend into the
diffuse \lya\ forest.  This was noted in the earliest simulation
studies of \lya\ absorption that explored outflows~\cite{the02}.
However, there are significant differences for stronger absorbers,
as outflows (particularly in the vzw case) deposit more cool gas
in the outskirts of halos, which increases the \lya\ absorbing
cross-section.  This yields significantly more sub-Lyman Limit
systems and slightly shallower CDD slopes, and it is also reflected
in the number counts of strong lines as we will show in \S\ref{sec:dndz}.

\subsection{$b$-parameter Distributions}\label{sec:bpar}

The right panels of Figure~\ref{fig:Nb} show the $b$-parameter
distributions for our three wind models at $z=0,1,2$.  The histograms
show the usual Gaussian distribution with an extended tail to higher
line widths.  The median $b$-parameter for each model is indicated by
the tickmark above the curves.  The typical value is around $40\kms$ at
all $z$, with a slight trend to become smaller at high-$z$.

The line widths are significantly wider than what was found using
STIS data~\citep[e.g.][]{dav01,wil10}.  Simulations by \citet{dav01}
predicted a median $b$ that depended on $\nhi$ with a typical value
of $25\kms$, in good agreement with their STIS data.  \citet{leh07}
determined a median $b=30\kms$, also from STIS data.  \citet{pas09}
used simulated spectra with approximately $5\kms$ resolution and
obtained a similar $b$-parameter distribution.  We will show in
\S\ref{sec:res} that the larger $b$ values predicted in this work
are mostly a result of the poorer spectral resolution of COS relative
to STIS.  Our simulated line widths are more comparable to those
obtained using GHRS \citep[$\sim 18\kms$ resolution;][]{pen04}.
Owing to this sensitivity to instrumental characteristics, we do
not conduct any detailed comparison with data here.  Nevertheless,
we will show in \S\ref{sec:phys} that line widths still contain
some information about the underlying temperature of the absorbing
gas.  It may be possible to extract better constraints on linewidths
using higher-order Lyman lines through a curve-of-growth analysis
or through simultaneous fitting of multiple
transitions~\citep[e.g.][]{leh07}, but we leave such a study for
the future.

The line widths, like the CDDs, show almost no sensitivity to outflow
model.  This is perhaps more surprising, given that the various wind
models heat the IGM to different levels (see Figure~\ref{fig:baryphase}),
with cw clearly providing substantial heating.  In \S\ref{sec:btemp},
we will show that the $b$ parameters are only loosely correlated with
temperature, reflecting the dominance of other sources of line broadening
such as Hubble flow and instrumental resolution.  In detail, the cw model
does have marginally higher median $b$ values of 43.0, 39.4, and $38.2\kms$
at $z=0,1,2$ respectively, as compared to vzw which has 42.3, 38.2, and
$36.2\kms$ (almost identical to nw), so the expected effect is present
but small.  We will also show in \S\ref{sec:res} that $b$ parameters are
highly sensitive to spectral resolution, and hence comparisons of data
to models (or to other data sets) must carefully account for such effects.

\subsection{Absorber Evolution}\label{sec:dndz}

\begin{figure}
\vskip -2.0in
\setlength{\epsfxsize}{0.6\textwidth}
\centerline{\epsfbox{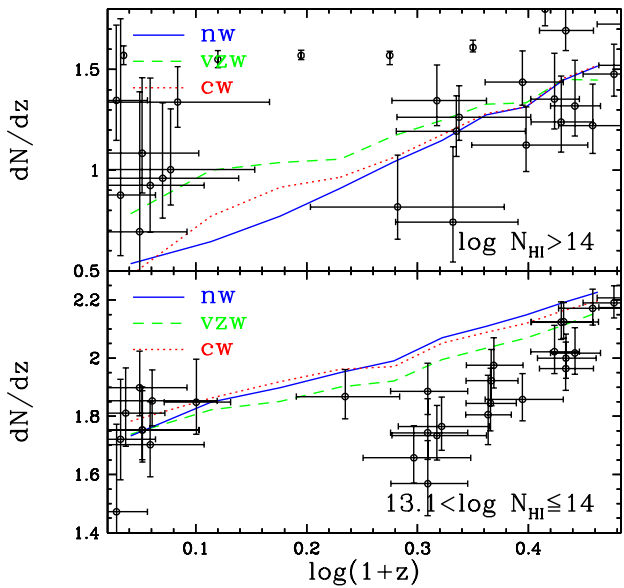}}
\vskip -0.5in
\caption{Line-of-sight number density of \lya\ absorbers ($dN/dz$), 
as a function of redshift.  Top panel shows the evolution of absorbers
with $\nhi>10^{14} \cdunits$, while bottom panel shows 
$10^{13.1}<\nhi\leq 10^{14} \cdunits$.  Data from various sources
compiled by \citet{wil10} are indicated.
}
\label{fig:dndz}
\end{figure}

The evolution of \lya\ absorbers provides insights into the interplay
between cosmic expansion, photoionization, and the growth of
structure.  One simple statistical characterization of this evolution
is the redshift-space number density of absorbers along the line
of sight, $dN/dz$.  

$dN/dz$ is usually measured within a specified column density range.
Early data from FOS \citep{wey98} showed a dramatic change in the
evolution rate of strong ($\nhi\ga 10^{14} \cdunits$) absorbers
between $z\ga 2$ and $z\la 1$.  However, subsequent higher-quality
data has muddled the situation.  Owing to the paucity of such strong
absorbers, there appears to be substantial cosmic variance for $dN/dz$
among the various sightlines.  Moreover, more recent data with
substantially higher resolution generally shows a significantly
lower $dN/dz$ compared to the FOS results, presumably because higher
resolution enables more accurate deblending of strong lines.  
It appears that large redshift path length and
high spectral resolution are both 
necessary to obtain an accurate estimate of $dN/dz$ for strong
absorbers; COS should provide this combination.

Figure~\ref{fig:dndz} shows the evolution of $dN/dz$ for our three
outflow models.  The top panel shows $dN/dz$ for $\nhi>10^{14}
\cdunits$ lines, while the bottom shows $10^{13.1}<\nhi\le 10^{14}
\cdunits$ absorbers.  Overlaid are a compilation of data taken from
\citet{wil10}.  For strong lines, the points across the top of the
panel are the Key Project FOS results~\citep{wey98}.  The others
come from higher-resolution data sets, and are all significantly
lower than the FOS results.  In general, the simulations show a
relatively slow decline in $dN/dz$, with little evidence for any
kind of ``break" in evolution as was seen in the FOS data.  
\cite{the98} and \cite{dav99} emphasized that the decline of the 
UV background at $z<2$ slows the evolution of the forest relative to
a constant UVB case.  The steadiness of evolution in Figure~\ref{fig:dndz}
relative to Figure~3 of \cite{dav99} is partly a consequence of
the narrower redshift range and partly a consequence of
adopting a UVB that declines more slowly at low redshift because
of contributions by galaxies.

The strong absorbers show a clear discrimination between outflow models
by $z=0$, with twice as many $\nhi>10^{14} \cdunits$ absorbers in 
the vzw model as in the no-wind case.  The cw model is intermediate.  
This shows
that our favoured outflow model is ejecting more gas into the vicinity
of galaxies where strong absorbers arise, and that this gas is remaining
relatively cool.  The difference arises because the momentum driven wind model
ejects more material from lower-mass galaxies at lower wind speeds,
thereby providing a reservoir of cooler gas that can cause strong \lya\
absorption.  This is qualitatively the same effect that causes this model
to provide a larger amount of DLA absorption and broader DLA kinematics,
as described by \citet{hon10}.  The vzw case shows the best agreement with
recent $z\sim 0$ data, which is directly tied to its somewhat shallower
CDD slope, although the uncertainties are still large enough to preclude
any firm discrimation. COS should yield a good observational determination,
providing an important diagnostic of galactic outflows.

The weaker absorbers show much less sensitivity to outflows, indicating
that winds do not substantially disturb gas at the densities giving
rise to $\nhi\la 10^{14}\cdunits$ absorbers.  A reasonable fit for all
the models is given by $dN/dz\propto (1+z)^{0.7}$, which is significantly
shallower than the evolution at high-$z$~\citep[e.g.][]{dav99}.  However,
all the models generically fail to reproduce the apparent break in evolution
at $z\sim 1$~\citep{jan06}.  This can be traced directly to the evolution
of the mean flux decrement $\bar{D}$ (Figure~\ref{fig:meanflux}), which
the models overpredict around $z\sim 1$ by the same factor of two that
they ovthe erpredict $dN/dz$.  In fact, the two observational data sets are
not unrelated, as $\bar{D}$ is most sensitive to absorbers just below
the logarithmic portion of the curve of growth, namely $\nhi\la 10^{14}
\cdunits$.
If we adjusted the UVB intensity to reproduce the \cite{kir07}
$\bar D$ measurements at $z=1-1.5$, then we would predict a break
in $dN/dz$ evolution at $\log(1+z) \approx 0.35$.
Improved statistics will thus provide valuable constraints on the 
evolution of the UVB, perhaps demonstrating significant departures
from the \citet{haa01} predictions.

In summary, the evolution of \lya\ absorbers will in principle
provide strong constraints on the evolution of the ionizing background
(for weaker lines), and perhaps on the nature of outflows (for
stronger lines).  However, accurate constraints
require larger data sets with good resolution,
which COS should provide.

\subsection{The Effects of Instrumental Resolution and Noise}\label{sec:res}

Intrumental resolution and noise characteristics of the data can have
a non-trivial impact on the derived statistics of \lya\ absorbers.
This can lead to some confusion when cross-comparing different samples,
or when comparing simulations to data without carefully matching such
characteristics.  In this section we use our simulations to briefly
investigate how COS-resolution data would compare to an equivalent sample
of higher STIS-resolution data, and how the \lya\ forest would appear
if it were possible to greatly boost the S/N of COS data.  These are
idealized experiments, not readily achievable with current instruments, but
they illustrate the trends one should keep in mind when intercomparing
samples.

\begin{figure}
\vskip -0.45in
\setlength{\epsfxsize}{0.65\textwidth}
\centerline{\epsfbox{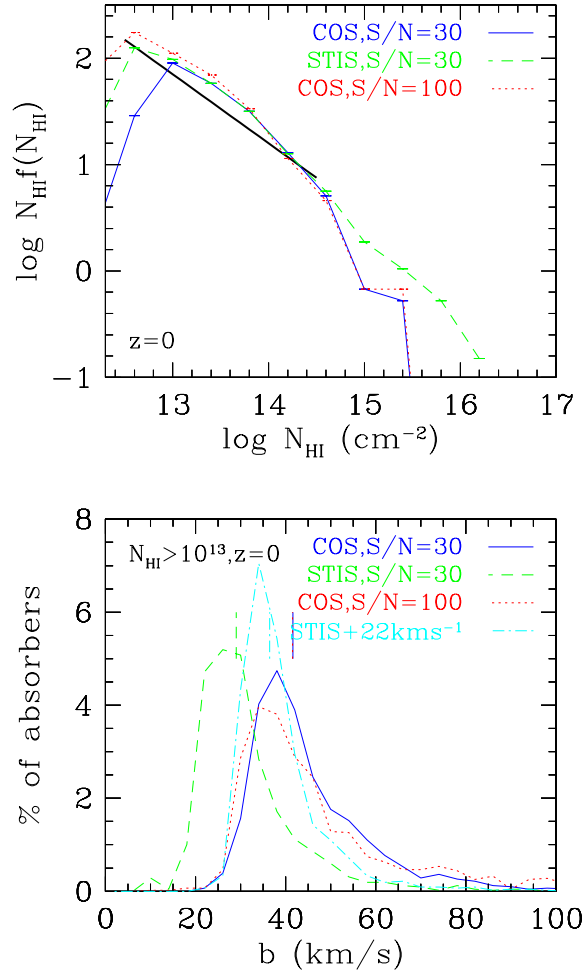}}
\vskip -0.5in

\caption{Effects of signal-to-noise ratio and spectral resolution
on Voigt profile parameter distributions in our fiducial vzw run at
$z=0$.  COS resolution assumes the COS LSF in the FUV channel, which is
$\approx 17\kms$ Gaussian width ($1\sigma$), 
while STIS resolution assumes a Gaussian with
$7\kms$ width.  {\it Top panel} shows column density distributions
multiplied by $\nhi$,
demonstrating that the turnover at low-$\nhi$ at $10^{13}\cdunits$ owes to
a lack of sensitivity; either increasing S/N or resolution allows one
to probe further.  There are also differences at the high-$\nhi$ end.
{\it Bottom panel} shows the $b$-parameter distributions; these are
not very sensitive to S/N variations, but are dramatically sensitive
to spectral resolution.  Here we also show the result of convolving the 
linewidths identified in the STIS-resolution spectra with a 22~km/s Gaussian
to illustrate the effect of moving from STIS to COS resolution; this
demonstrates that much of the line broadening in the COS-resolution case 
owes to its coarser spectral resolution.
}
\label{fig:Nbres}
\end{figure}

Figure~\ref{fig:Nbres} shows the CDD and $b$-parameter distributions for
the vzw model at $z\approx 0$ for our simulated spectra convolved with
a COS LSF at S/N=30 and 100, and with a $7\kms$ Gaussian (with
$3\kms$ pixels) intended to roughly mimic STIS resolution.  The higher
S/N or resolution both extend the CDDs to lower column densities with an
unbroken power law, showing that our COS-resolution sample with S/N=30
is essentially complete to $\nhi=10^{13} \cdunits$.  Coincidentally,
increasing the S/N to 100 at COS resolution is roughly equivalent,
in terms of CDD completeness, to having STIS resolution with S/N=30.
At the high-$\nhi$ end, STIS resolution produces quite a few more strong
lines.  We caution that different Voigt profile fitting algorithms can
significantly affect the parameters derived for saturated lines, so
one must be careful in interpreting this plot.  However, it does
indicate that some of the variations seen in $dN/dz$ for strong lines
may owe to different instrumental characteristics for the various data
sets.  
Examination of the true column densities in the simulated absorbers
shows that the high column density systems identified in the STIS-resolution
spectra are generally real; at COS resolution \autovp\ mischaracterizes
them as lower column density systems with larger $b$-parameters
(and thus less saturation).
Finally, the slope of the CDD  at $10^{13}<\nhi<10^{14.3}\cdunits$
is steeper in the case of higher S/N or resolution; for the former,
we obtain a slope of $-1.73$ (as opposed to $-1.70$ for the original
vzw case), and for the latter we get $-1.82$.  Hence, high resolution
produces a significantly steeper CDD slope, which can help explain why
the CDD slopes in \citet{dav01b} are steeper.

The $b$-parameter distribution shows strong sensitivity to spectral
resolution, but increasing S/N does not have a major impact on $b$
parameters.  At STIS resolution, the median $b$-parameter is below
$30\kms$, and the distribution roughly agrees with that found in
\citet{dav01}.  At COS resolution, lines are significantly broader,
resulting in a median greater than $40\kms$ and a much more prominent tail
to high $b$ values.  The cyan dot-dashed curve in Figure~\ref{fig:Nbres}
shows the result of taking each line in the STIS-resolution distribution
and increasing its $b$-parameter by adding $\sqrt{2(17^2-7^2)} \approx
22\kms$ in quadrature to account for the difference in resolution
between COS and STIS.  The peak of the distribution moves much closer
to that of the COS distributions, suggesting that for many lines the
larger COS $b$-parameter is a simple consequence of convolving the
line with a broader LSF.  However, the COS distributions still have
many more lines at $b\geq 60\kms$, showing that these broad lines must
arise mainly from blends that cannot be resolved by the COS FUV channel.
Because of the effects of line-blending, we think it is generally better
to test models by applying the LSF to simulated spectra rather than
attempting to correct measured $b$-parameters on a line-by-line basis.
The exception might be in cases where higher order Lyman series lines
provide additional constraints on the $b$-parameters via curve-of-growth
analysis.  More generally, this illustrates that the $b$ parameters
are particularly sensitive to spectral resolution, and any comparisons
between data sets, or between data and models, should take special care
to match the instrumental line spread function.

\section{\lya\ Absorber Physical Conditions}\label{sec:phys}

We now examine the relationship between absorber properties and the
physical conditions of the absorbing gas, namely its temperature,
density, and ionization state.  The relatively simple nature of
\lya\ forest absorption, particularly for weaker absorbers, results
in fairly tight relationships that offer insights into the evolution
of baryons in the IGM.

To assign densities and temperatures to individual absorbers, we
find the location of peak absorption and assign to the absorber the
\ion{H}{i}-weighted $\rho$ and $T$ of the nearest pixel.  This is not a
perfect procedure, as, e.g., the width of the line might be determined by
coincident hotter gas superposed on stronger absorption from colder gas.
But it gives a reasonable idea of the physical conditions present in the
absorbing gas.  To better isolate real trends, we only consider absorbers
whose formal $\chi^2$ errors for $\nhi$ and $b$ as reported by \autovp\ are
fractionally less than 20\%, which mostly removes very weak absorbers
whose parameters are not well-determined.  For studying linewidths, we
include only absorbers with $\nhi>10^{13}\cdunits$, for which we have
a relatively complete sample.  We focus on the momentum-driven wind
case and point out where substantial differences arise in the other
wind models.

\subsection{Absorbers in Phase Space}

\begin{figure*}
  \subfigure{\setlength{\epsfxsize}{0.49\textwidth}\epsfbox{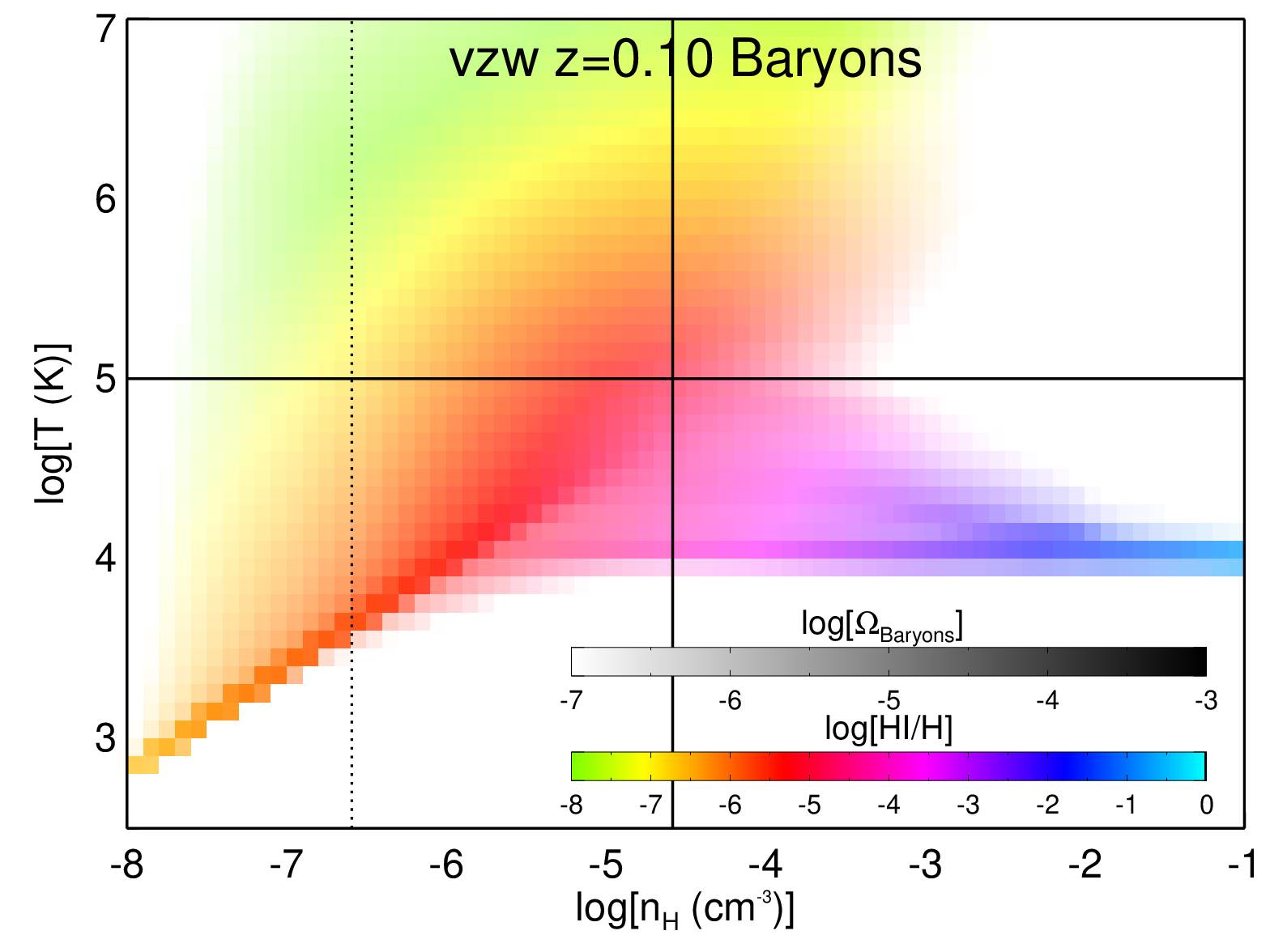}}
  \subfigure{\setlength{\epsfxsize}{0.49\textwidth}\epsfbox{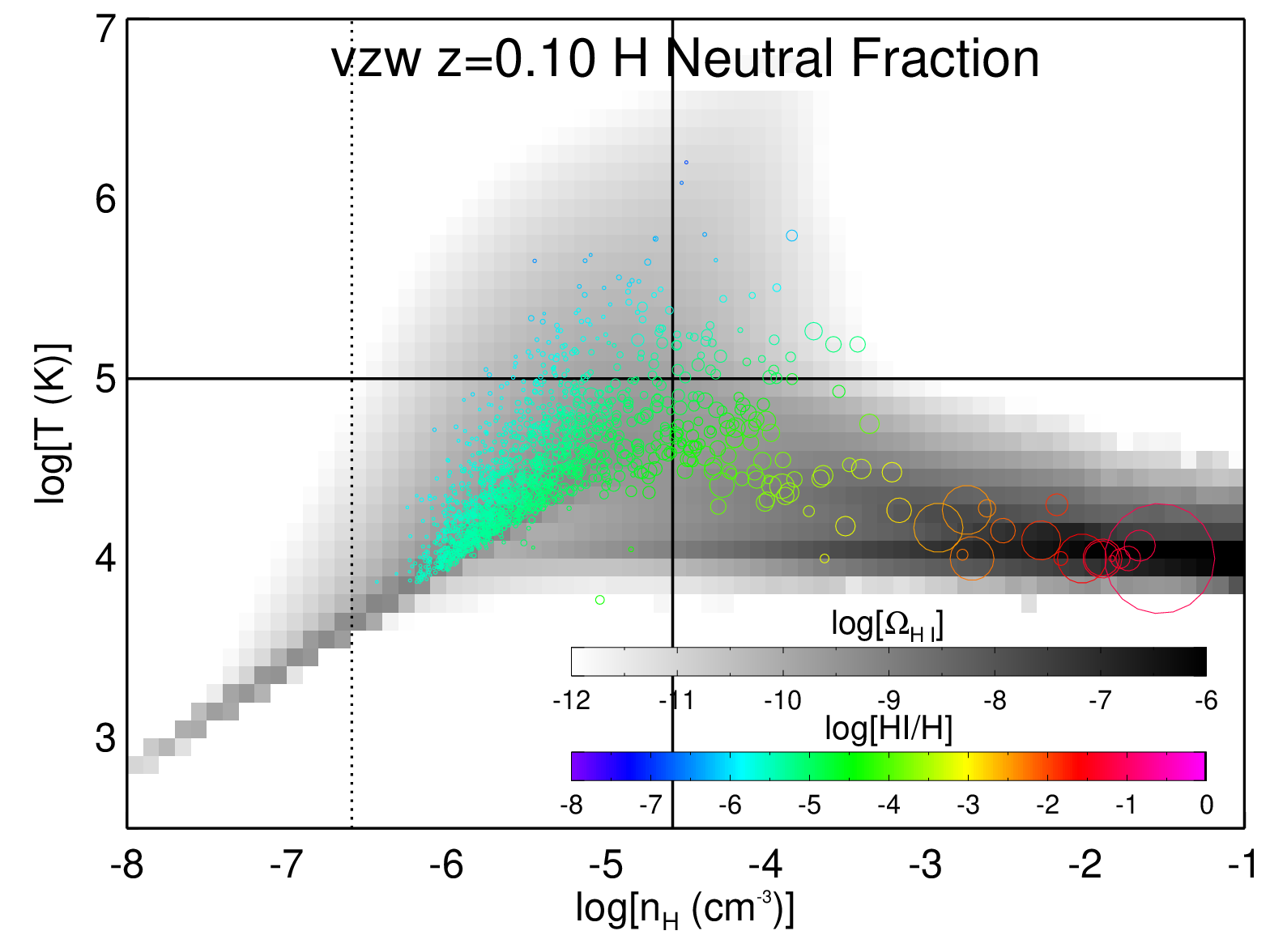}}
  \subfigure{\setlength{\epsfxsize}{0.49\textwidth}\epsfbox{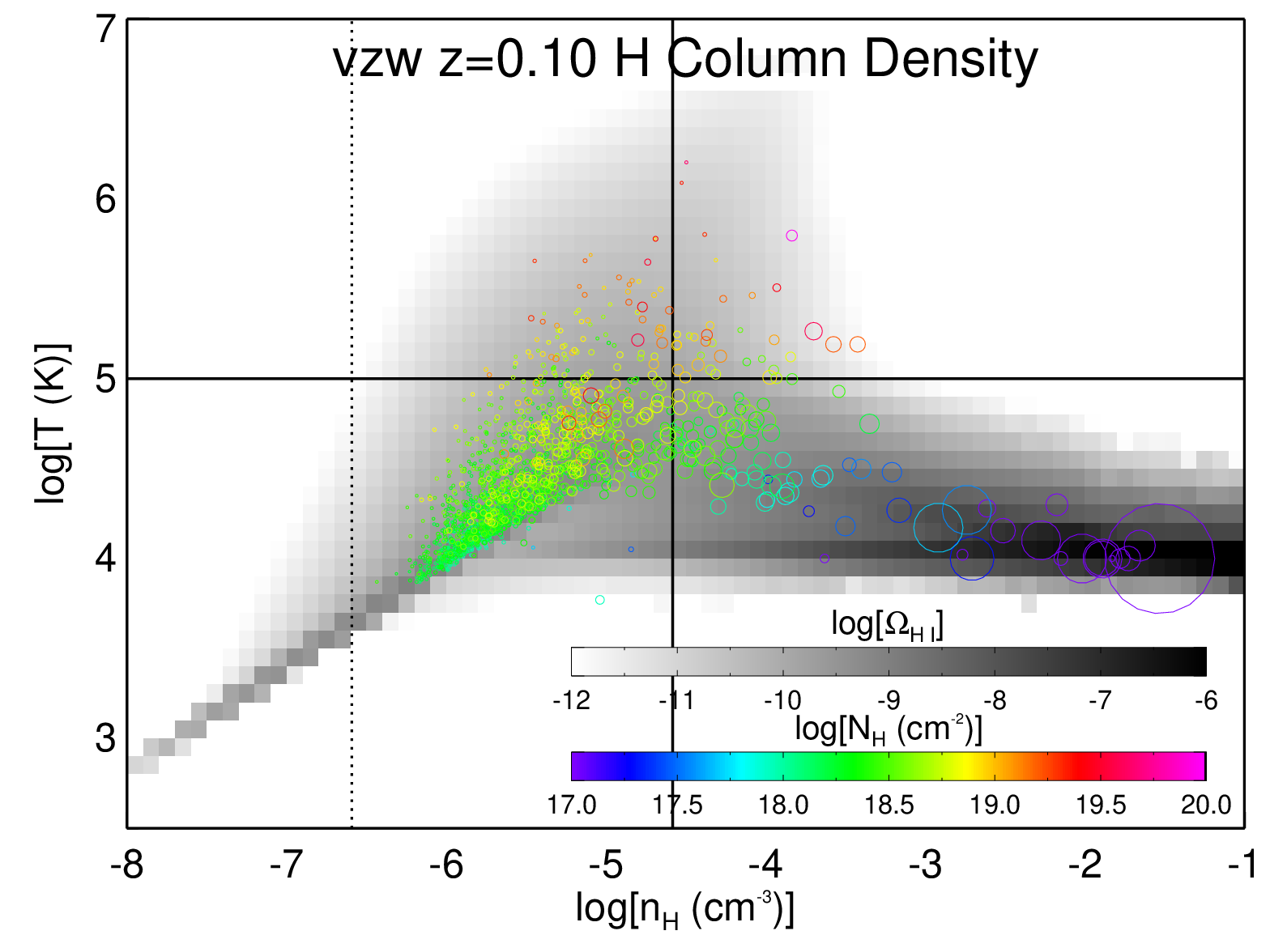}}
  \subfigure{\setlength{\epsfxsize}{0.49\textwidth}\epsfbox{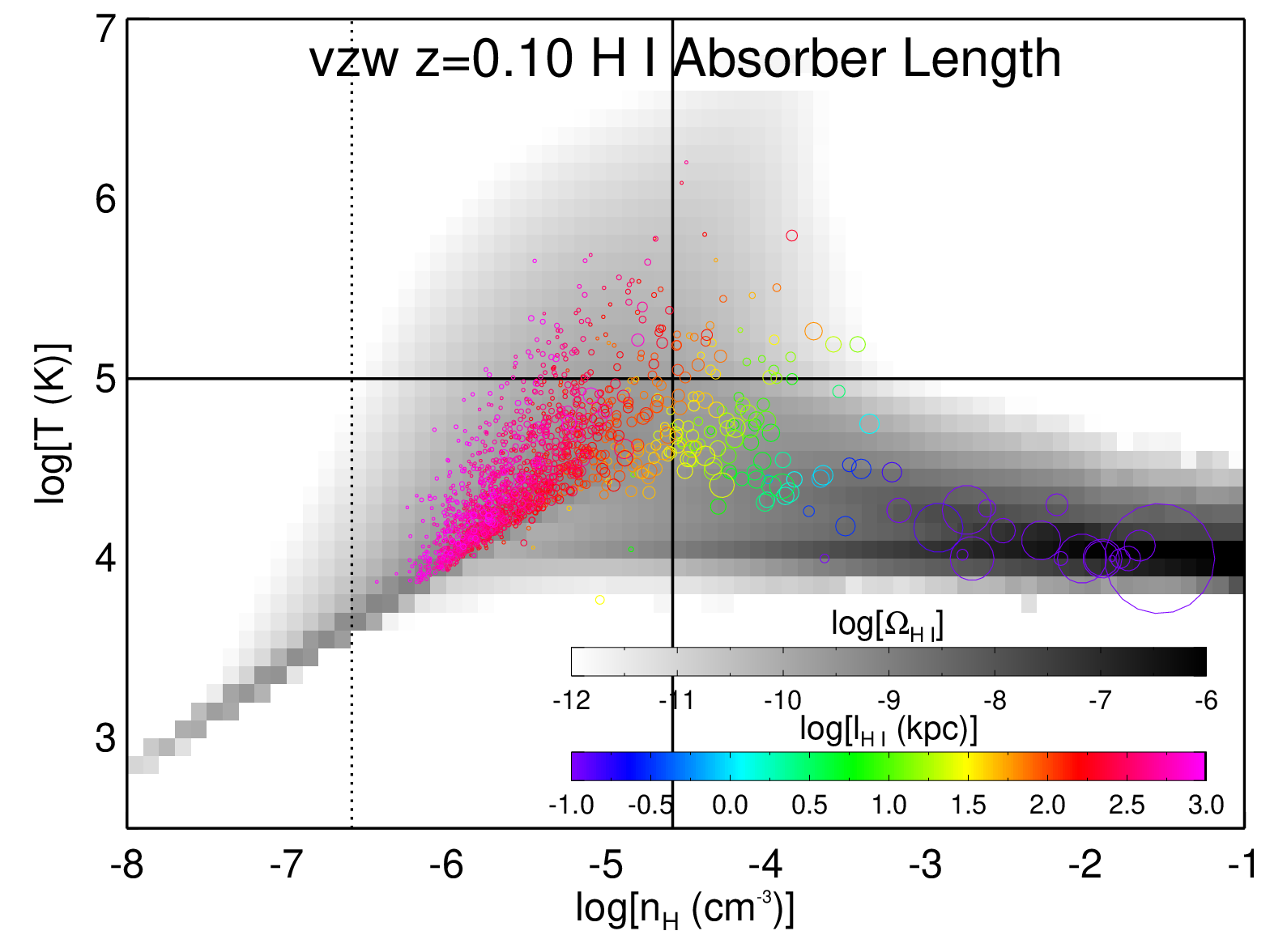}}
  \subfigure{\setlength{\epsfxsize}{0.49\textwidth}\epsfbox{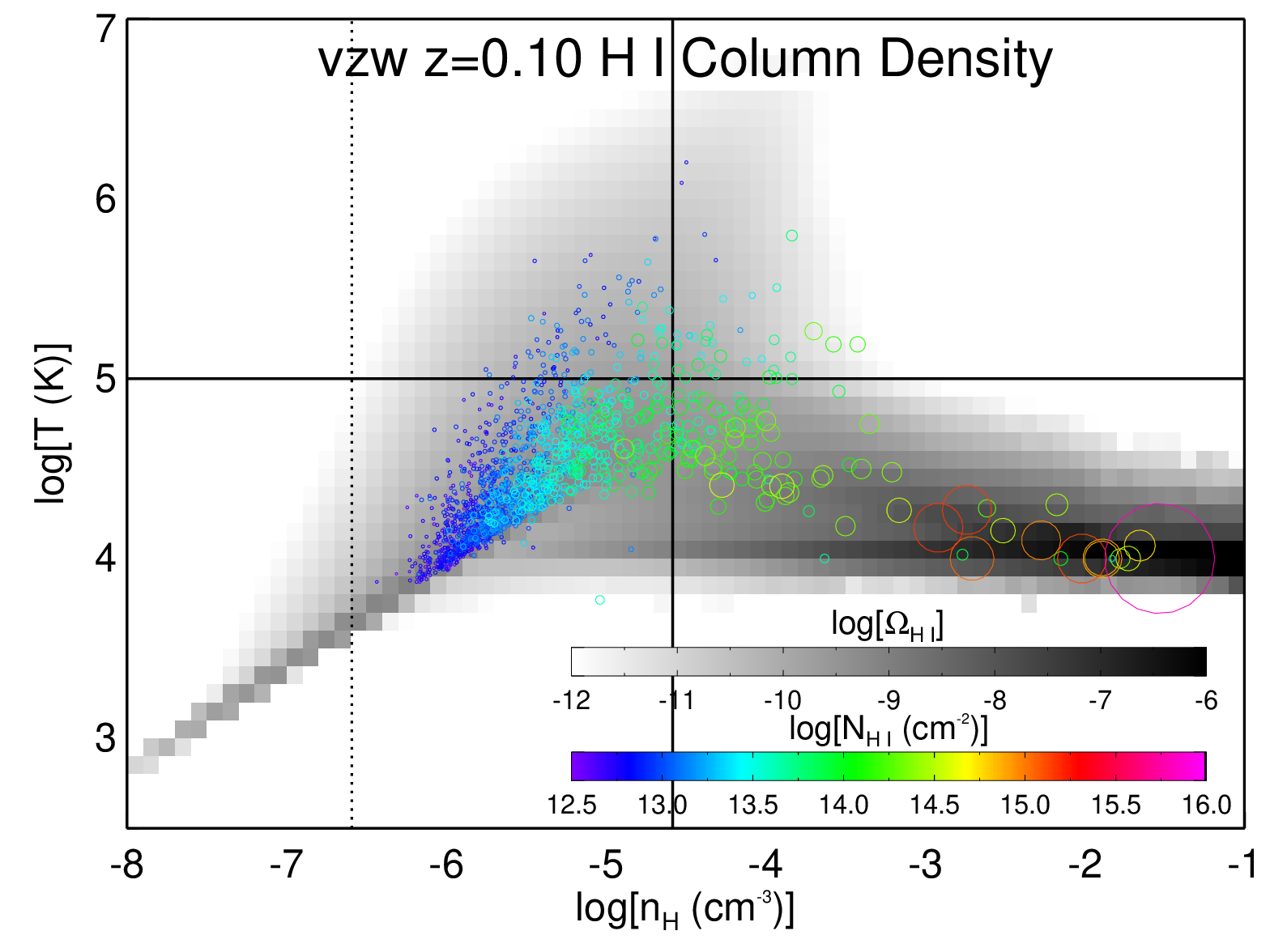}}
  \subfigure{\setlength{\epsfxsize}{0.49\textwidth}\epsfbox{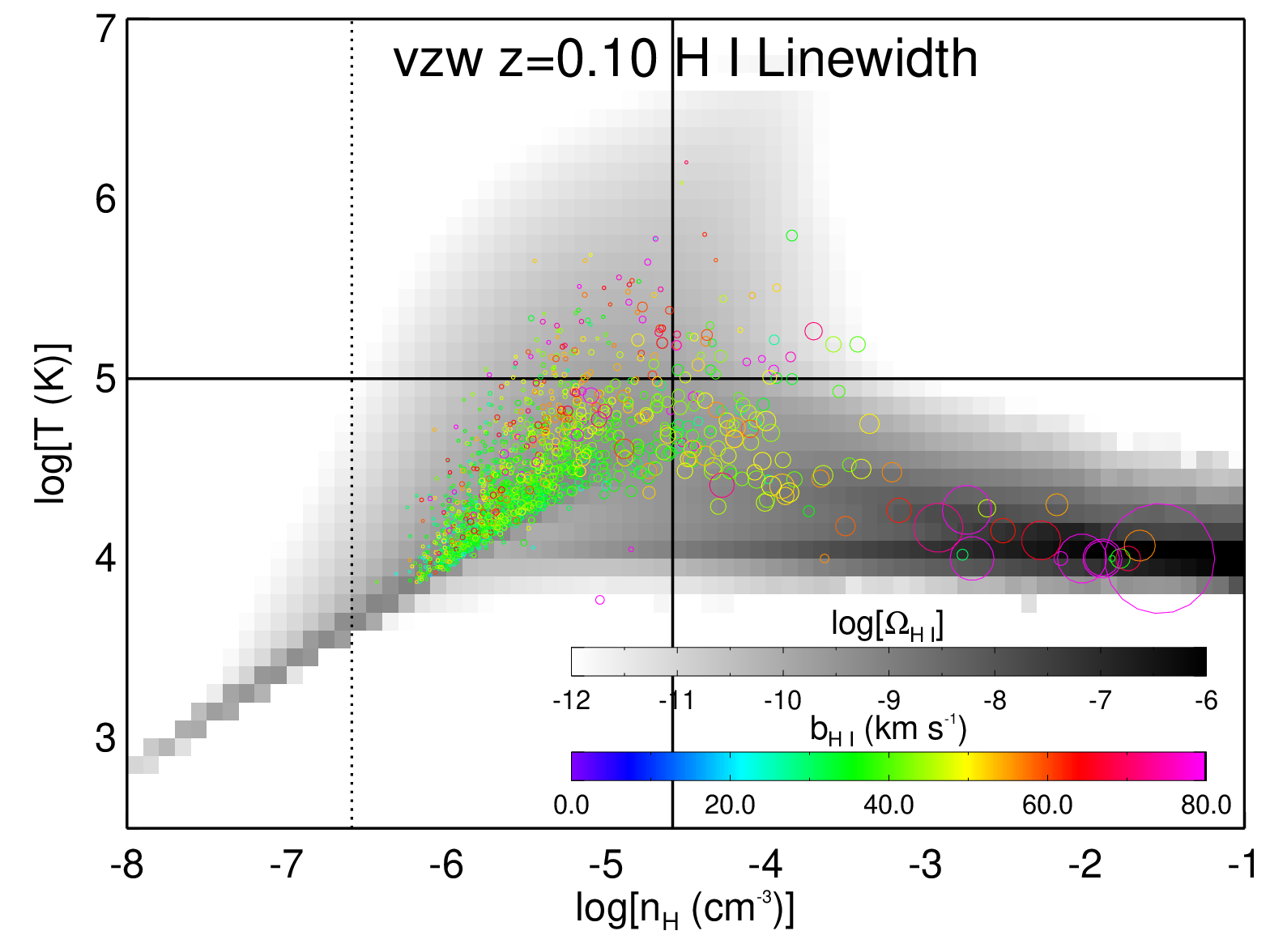}}
\caption{Cosmic phase diagrams for the vzw run at $z=0.1$.  Upper left
shows all baryons, colored by \ion{H}{i} fraction.  Other panels
are shaded by $\Omega_{HI}$, binned in $0.1\times 0.1$~dex pixels.
Solid lines demarcate the four cosmic phases as in Figure~\ref{fig:rhoT},
while the dotted line indicates the cosmic mean density.  In each panel,
the circles show the locations of $z=0-0.2$ absorbers in phase space,
with the area of the circle proportional to $\nhi$ (see lower left
panel for scale).  The color scale
in each panel corresponds to different quantities: Ionization fraction
(upper right), total hydrogen column density $N_H$ (middle left), absorber
size (middle right), $\nhi$ (lower left) and linewidth (lower right).
}
\label{fig:phasespace}
\end{figure*}

Figure~\ref{fig:phasespace}
shows how \lya\ absorbers populate and trace cosmic
phase space (here shown as temperature vs. hydrogen number density).
In each panel except the
upper left, our simulated absorbers between $z=0-0.2$ are indicated
as circles, placed using their \ion{H}{i}-weighted densities and
temperatures.  Solid lines demarcate the phase boundaries as defined
in \S\ref{sec:phases} and
shown in Figure~\ref{fig:rhoT}, and the vertical dotted line indicates
the cosmic mean density of hydrogen.  Our absorber sample overlaps the
shading, which represents the simulated distribution of \ion{H}{i}.
However, because random sight lines sample the \lya\ forest in a
volume-weighted fashion, there are more weak \lya\ absorbers tracing
lower densities even though most of the cosmic \ion{H}{i} resides in rare,
strong absorbers tracing the condensed phase.  Conversely, absorbers
are not detectable at densities below $n_H = 10^{-6}$~cm$^{-3}$, where
the vast majority of the IGM volume exists, and 
this detectability threshold is the reason that the low-z \lya\
forest is observed to be sparsely populated.

The upper left panel displays the cosmic distribution of baryons in
phase space, color-coded by ionization fraction.  The remaining 5
plots contain up to three additional dimensions of information within
two-dimensional cosmic phase space: Shading for global \ion{H}{i} content
(i.e. the product of the intensity and color maps in the top left panel),
circle size scaled to absorber $\nhi$, and circle color for the various
quantities, as follows: Neutral fraction (upper right), total hydrogen
column (middle left), absorber path length (middle right), $\nhi$ (lower
left), and linewidth (lower right).  It is clear that there are strong
correlations amongst the various physical and observational quantities.
In subsequent sections we will quantify some of these trends in more
detail and attempt to connect observables to physical quantities within
this plane.

The \ion{H}{i} neutral fraction in the upper right panel is computed for
each absorber as an optical depth-weighted neutral fraction from all
SPH particles that contribute to the absorption.  Note that we assume
ionization equilibrium; this is expected to be a very good approximation
in the \lya\ forest.  It is clear that, as expected, the IGM is highly
ionized, with the weakest absorbers showing neutral fractions of $\sim
10^{-6}$~cm$^{-3}$.  Neutral fractions increase towards lower temperature and
higher density, such that absorbers at $\sim 10^{14}\cdunits$ (the
typical absorber at the crossing point of the phase division lines)
show a neutral fraction of $\sim 10^{-5}$~cm$^{-3}$.  The neutral fraction
increases rapidly along the condensed phase, leading quickly to
very strong absorbers.

The middle left panel of Figure~\ref{fig:phasespace} shows the total
hydrogen column density $N_H$, computed by dividing $\nhi$ by the
neutral fraction.  A remarkable trend emerges that the vast majority
of \lya\ forest absorbers have $N_H$ between $10^{18}\cdunits$ and
$10^{19}\cdunits$, except for the hottest or densest absorbers.  This
trend was noted observationally in a small sample by \citet{pro04}, which
they pointed out was in general agreement with simulation predictions
from \citet{dav01}.  Here the trends of $N_H$ versus overdensity can be
more clearly seen; it is double-valued in the sense that $N_H$ is lower at
both low and high overdensities, and it peaks at intermediate overdensities,
particularly at high temperatures.

The middle right panel shows an estimated size of \lya\ absorbers,
obtained by dividing $N_H$ by the physical density.  Weak absorbers ($\sim
10^{13}\cdunits$) have sizes up to $\sim 1$~Mpc, and this quickly
drops to $\sim 100$~kpc for $\sim 10^{14}\cdunits$.  This is
broadly consistent with the large coherence lengths observed in quasar
pair sightlines~\citep[e.g.][]{din98,cas08}, 
though the elongated geometry of absorbing structures
also affects apparent coherence lengths.
The sizes are consistent with analytic estimates by \citet{sch01}
from assuming that Jeans smoothing establishes the structure of the
low-density IGM.  
(See \citeauthor{pee10} [\citeyear{pee10}] for evaluation of this
hypothesis against simulation results, albeit at higher redshift.)
The absorber lengths drop rapidly with increasing
overdensity, such that strongest absorbers in the condensed phase have
sizes $\sim 1$~kpc or less.  This explains why absorbers from this
phase are rarely seen despite the substantial reservoir of neutral gas
in the condensed phase --- their cross-section is remarkably small.

The two lower panels consider absorber observables $\nhi$
and $b$.  
The lowest column density systems lie along the low-density,
high-temperature envelope of detectable \ion{H}{i}.  The trend of
increasing $\nhi$ towards low temperatures and high densities mimics
that in the ionization fraction, such that the typical absorber at
the bound/unbound division has $\nhi\sim 10^{14}\cdunits$ (and lower
at higher temperatures).  The strongest absorbers generally arise in
condensed phase gas, though they are rare in these random LOS.

The $b$-parameters in the lower right panel also show trends in phase
space, but they are not as distinct as with column densities.  Narrower
lines tend to arise in lower-density and lower-temperature gas, but there
is a significant scatter. Wide lines can occur in hot gas, but such
gas can also contain narrower lines.  Wide lines can also occur in very
strong absorbers that are heavily saturated, arising in condensed gas.
Overall, $b$ parameters trace temperatures fairly loosely, a point we will
quantify further in \S\ref{sec:btemp}.

Overall, \lya\ absorption along random lines of sight can trace
present-day baryons from close to the mean overdensity up to overdensities
of $\sim 10^4$ or more, and to temperatures well above $10^5$~K.
There are strong trends of ionization fraction and absorber size, which
together with the underlying baryon distribution in phase space yield a
strong trend of physical parameters versus column density, and a weaker
but still noticeable trend versus linewidths.  In the next two sections
we explore these trends more quantitatively, to better understand how
observables obtained from COS spectra can reveal the physical conditions
of the absorbing gas.

\subsection{Column density versus physical density}

\begin{figure}
\vskip -0.5in
\setlength{\epsfxsize}{0.6\textwidth}
\centerline{\epsfbox{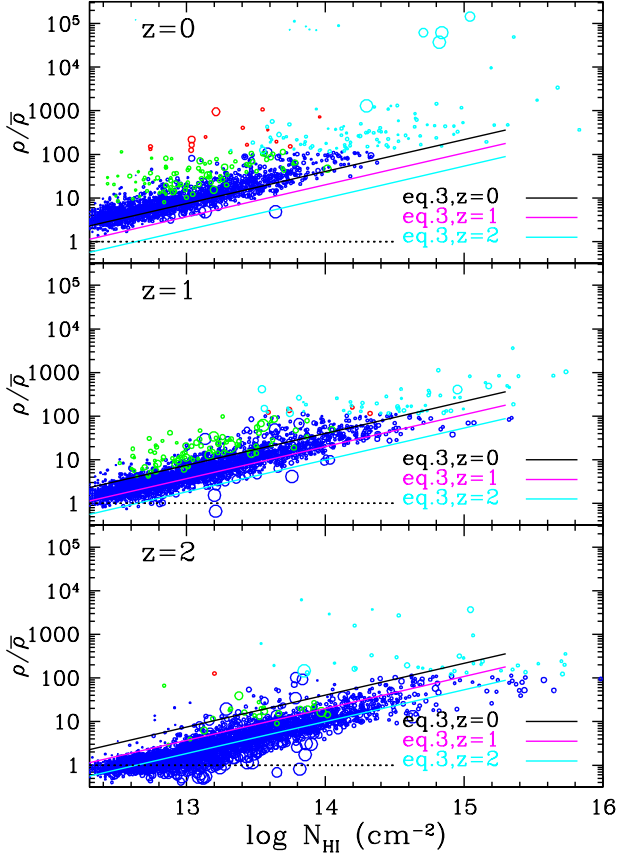}}
\vskip -0.5in
\caption{Baryonic physical density in units of the cosmic mean
($\rho/\bar\rho$) versus column density $\nhi$ for absorbers from
the momentum-driven wind model, at $z=0$ (top), $z=1$ (middle), and
$z=2$ (bottom).  In each panel, the solid line shows the relation from
equation~\ref{eqn:NHIdelta} at $z=0$ (black), $z=1$ (magenta) and
$z=2$ (cyan).  Absorbers are color-coded as diffuse
(blue), WHIM (green), hot (red), or condensed (cyan).
Point sizes are scaled $\propto b$.
}
\label{fig:NHIdelta}
\end{figure}

The evolution in the relationship between column density $\nhi$
and overdensity $\delta$ is governed by the interplay between
cosmic expansion, 
which lowers the physical density (at a given $\delta$) and thus
lowers the recombination rate, and evolution of the UV background,
which drops at lower redshifts and thus lowers the photoionization
rate.
\citet{dav99} showed that much of the observed evolution in the \lya\
forest can be understood through the evolution in this relation, because
this interplay causes a given strength absorber to correspond to a higher
overdensity at lower redshift.  Complications arise because, by $z=0$,
a significant amount of IGM gas is heated collisionally by shocks rather
than by photoionization, resulting in departures from the FGPA~\citep{zha05}.
Nevertheless, this $\nhi-\delta$ relation is central to understanding the
physics of the \lya\ forest in a structure formation context.

Figure~\ref{fig:NHIdelta} shows that a fairly tight $\nhi-\delta$ relation
is established at all redshifts, as found by previous studies.  The three
panels show this relation for the momentum-driven wind model at $z=0, 1, 2$.
Absorbers are categorized by cosmic phase, with a color code following
Figure~\ref{fig:baryphase}, namely diffuse (blue), WHIM (green), hot halo
(red), and condensed (cyan).  The lower envelope reflects gas that is
purely photoionized, as shock heating will move absorbers to lower $\nhi$
for a given overdensity by lowering the neutral fraction.

The diffuse-phase absorbers follow a tight relation in overdensity
versus temperature owing to photoionization heating and cosmic
expansion~\citep{hui97,sch01}.  We perform a least-squares fit to
the diffuse absorbers as a function of column density and redshift
(from $z=0-2$), restricting ourselves to absorbers with $T<10^{4.5}$~K
to remove mildly shocked systems, and obtain the following relation:
\begin{equation}\label{eqn:NHIdelta}
\frac{\rho}{\bar\rho} = \frac{35.5\pm 0.3}{f_\tau^{0.741}} \Bigl(\frac{\nhi}{10^{14}\cdunits}\Bigr)^{0.741\pm0.003} 10^{(-0.365\pm0.009)z}.
\end{equation}
Recall that $f_\tau$ is the factor by which optical depths have been
multiplied
relative to those computed for the \citet{haa01} ionizing background, 
to match the mean flux decrement; it is $2/3$ for our momentum-driven
wind model.  The quoted uncertainties are on means, and they do not
reflect the spread amongst the individual absorbers.
This fit is shown at each redshift in each panel, to help visualize
the rate of evolution.  Our slope for $\rho$ vs. $\nhi$ is similar 
to the slope of two-thirds derived by \citet{sch01} from Jeans smoothing 
arguments.  Furthermore, the redshift evolution is in excellent agreement
with predictions from his analytic model.

The WHIM absorbers generally have somewhat higher density at a given
$\nhi$, since the gas has been shock-heated; this trend is exacerbated
for the hot halo absorbers.  As a result, at overdensities above 10 or so,
corresponding to $\nhi\ga 10^{14}\cdunits$ at $z\sim 0$, there begin to be
significant departures from equation~(\ref{eqn:NHIdelta}).  The departures
from this relation also become more prominent with time, since more IGM
gas becomes shock-heated.  Condensed phase gas typically gives rise to the
strongest \lya\ absorption systems seen at the highest overdensities.
At $z=2$ there is a shelf at the lowest densities,
largely because the underdense absorbers have larger velocity widths
and are thus more difficult to identify at fixed $\nhi$ and S/N.

Equation~(\ref{eqn:NHIdelta}) is similar in form to the (by-eye) fits
presented in \citet{dav99} and \citet{dav01}.  Differences arise because
the ionizing background strength employed here is slightly different,
and here we consider only the diffuse absorbers.  We have checked that
the above equation holds reasonably well for the diffuse absorbers in the
no wind and constant wind cases (when the $f_\tau$ factor is included),
since these absorbers are essentially unaffected by outflows.  The other
wind models are qualitatively similar in terms of their other phases
as well, though in detail there are minor differences that we will discuss in
the following sections.  The insensitivity to winds further emphasizes
the fundamental nature of the $\rho-\nhi$ relationship in describing
the physics of the \lya\ forest.

\subsection{Absorption phase}\label{sec:absphase}

\begin{figure}
\vskip -0.2in
\setlength{\epsfxsize}{0.5\textwidth}
\centerline{\epsfbox{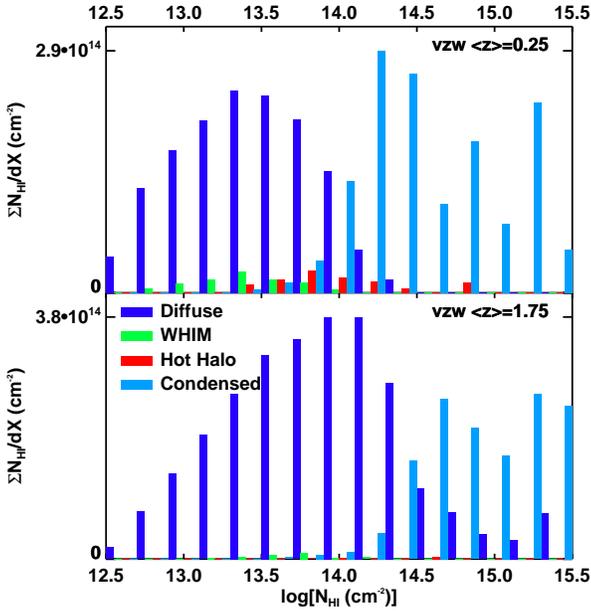}}
\vskip -0.1in
\caption{Summed column densities per unit path length as a function of
column density, subdivided by absorber phase.  The top panel shows absorbers
in the range $z=0-0.5$, and the bottom panel shows $z=1.5-2$.
}
\label{fig:absphase}
\end{figure}

We now examine the cosmic phase of \lya\ absorption as a function of
column density.  Figure~\ref{fig:absphase} shows a sum over \ion{H}{i}
column density for lines in bins of $\nhi$ for our momentum-driven
wind model.  At both low ($\bar{z}=0.25$) and high ($\bar{z}=1.75$)
redshifts, there is a 
clear transition in which total \ion{H}{i} absorption is dominated
by diffuse absorbers at low $\nhi$ and by condensed absorbers
at high $\nhi$.  The transition occurs at $\nhi\approx 10^{14}\cdunits$
for $\bar{z}=0.25$ and $\approx 10^{14.5}\cdunits$ for $\bar{z}=1.75$.
WHIM and hot halo absorbers are sub-dominant at all column densities,
though they are much more common at $\bar{z}=0.25$ than at $\bar{z}=1.75$.
While the column density demarcation is less clear for these phases
than for diffuse/condensed, WHIM absorbers typically have
$\nhi < 10^{14}\cdunits$, while hot halo absorbers sometimes have
larger $\nhi$.

While the broad trends are similar for other wind models, in detail there
are substantial differences particularly for absorption in bound phases.
If we consider absorbers from $10^{14}\la \nhi\la 10^{15.5}\cdunits$
at $z=0-0.5$, then the cw model contains only half the absorption, and nw
only $\approx 40\%$ of the absorption, relative to vzw.  This reiterates
the result in Figure~\ref{fig:dndz} that high-$\nhi$ absorbers provide
the most sensitive probe of galactic outflows in the \lya\ forest.

In summary, the low-$z$ \lya\ forest contains absorbers from all cosmic
phases, but diffuse absorbers dominate at $\nhi\la 10^{14} \cdunits$
and condensed absorbers dominate at higher $\nhi$.  WHIM and hot
phase absorbers are present, but highly subdominant in total absorption.
Next, we assess whether linewidths add enough information 
to pinpoint the absorbers arising in WHIM and hot halo gas.

\subsection{Linewidths and temperatures}\label{sec:btemp}

In the absence of other sources of line broadening, the $b$-parameters
will reflect the temperature of the underlying absorbing gas.  A
particularly promising application of this is tracing the missing
baryons with wide \lya\ lines because, unlike relying on high-ionization
metal lines such as \ion{O}{vi}~\citep[e.g.][]{tri00} or
\ion{O}{vii}~\citep[e.g.][]{nic05}, \ion{H}{i} linewidths are
independent of metallicity.  However, in practice, linewidths are
not dominated by thermal broadening~\citep{wei97c}, although the
thermal contribution is higher at low-$z$~\citep{dav01b}.  Here we
investigate the relation between linewidths and temperature in our
simulations, to better understand how robustly one can trace the
WHIM, and hotter IGM gas in general, using wide \lya\ lines.

\begin{figure}
\vskip -0.4in
\setlength{\epsfxsize}{0.65\textwidth}
\centerline{\epsfbox{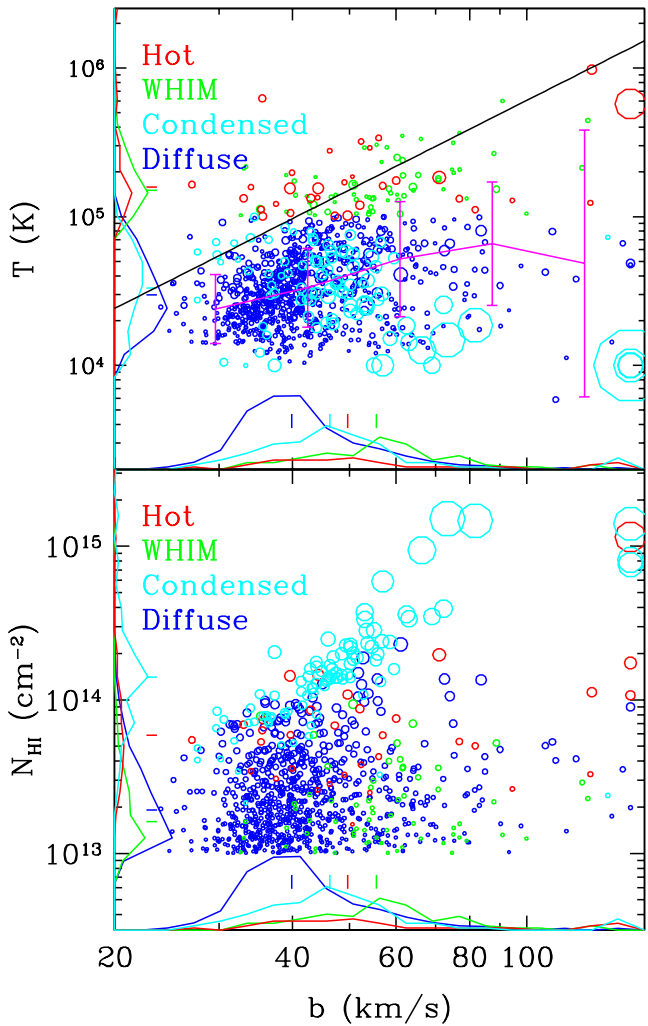}}
\vskip -0.2in
\caption{
{\it (Top)} Temperature vs.
$b$-parameter for \lya\ absorbers, color-coded by
absorber phase, from the vzw simulation at $z=0$.
Symbol size is scaled by $\sqrt{\nhi/10^{14}}$.
Diagonal solid line shows the relation for pure thermal
broadening; some absorbers are narrower than ``allowed" owing to
uncertainties in Voigt profile fitting and the assigning of temperatures
to individual absorbers.  Histograms in $b$ and $T$ are shown along the bottom
and side, respectively, divided by absorber phase, with tick marks above 
the curves identifying the median values.  For visibility of the
other phases, the diffuse-phase histograms have been reduced by $5\times$;
otherwise, it would dominate at essentially all $b$ values.
Magenta curve shows a running
median, with $1\sigma$ dispersion, for the entire sample.
{\it (Bottom)} Like top, but showing \ion{H}{i} column density in
place of temperature.
}
\label{fig:btemp}
\end{figure}

Figure~\ref{fig:btemp}, top panel, shows the distribution of
absorbers' $b$-parameters versus temperature from our momentum-driven
wind simulation at $z\approx 0$ (other wind models are similar).
Absorbers are color-coded by cosmic phase as in Figure~\ref{fig:baryphase}.
Recall that we do not fit a continuum to the simulated spectra,
hence there is no additional uncertainty in this regard in the
identification of broad absorbers in the models.  However, such
uncertainties may be important when analysing observations, as broad
absorbers may be partially ``fit out" when fitting a continuum to
unnormalized data.  There is a definite trend for wider absorbers
to arise in hotter gas, as shown by the running median (magenta
curve), but there is also substantial scatter as indicated by error
bars which show $1\sigma$ dispersion values about the median.  Hence,
a wide \lya\ line is more likely to arise in hotter gas, but it
does not always do so.

If lines were purely thermally broadened, then the \ion{H}{i} linewidth
would be related to the temperature by $b=12.8 \sqrt{T/10^4 {\rm K}}\kms$.
This relation is shown as the solid diagonal line.  A small fraction of
WHIM and hot absorbers are narrower than allowed by thermal broadening,
either because of $b$-parameter uncertainties in the Voigt profile
fits or because of uncertainties in assigning temperatures to
absorbers as described above.  Generally, however, most absorbers'
widths significantly exceed the thermal broadening value.  The running
median shows a similar trend as the thermal broadening track, but it
is higher by a factor of $1.5-3$ in $b$, until a turnover at the very
largest $b$ values.  For the narrow diffuse absorbers, we obtain the
same result as in \citet{dav01b} that thermal broadening contributes equally
to other sources (in quadrature) in setting the linewidths.  For wider
absorbers, the thermal contribution becomes smaller.  In general, the
dominant contribution to the $b$-parameters is not thermal broadening,
but rather Hubble broadening in the expanding gas or line blending owing
to limited spectral resolution as discussed in \S\ref{sec:res}.

Along the bottom and side of Figure~\ref{fig:btemp} are displayed absorber
histograms in $b$ and $T$ for each cosmic phase, with the median values
indicated by tick marks.  The diffuse-phase histogram has been multiplied
by $1/5$ along each axis for visibility.  The ones along the temperature
axis show the expected behavior, with hotter phases having higher median
temperatures.  The increasing $T$ with $b$ is broadly reflected in the
histograms along the $b$-axis as well, with WHIM and hot absobers being
at higher $b$ values than diffuse and condensed ones.  However, the scatter
between $b$ and $T$ is quite significant; the histograms of the various phases
span a wide range in $b$ and have a substantial  overlap.  Moreover, the
diffuse absorbers essentially dominate at all $b$ values (remember that this
histogram has lowered by $5\times$) simply because they are so numerous,
despite the fact that their median linewidth is substantially smaller than
all the other phases.  Hence, it is unfortunately not straightforward to
assign a gas temperature purely from the $b$ parameter.  A key implication
is that searches for WHIM gas with COS by selecting only broad \lya\
lines is likely to pick up a significant fraction of diffuse absorbers,
and care must be taken to avoid double-counting baryons in such studies.

Both the column density and linewidth provide some guidance in determining the
cosmic phase of the absorbing gas.  It stands to reason then that
combining the two observables may provide more accuracy in identifying
the cosmic phase.  The bottom panel of Figure~\ref{fig:btemp} shows
the absorbers in $\nhi-b$, color-coded by absorber phase.  Broad, weak
absorbers with $b>60\kms$ and $\nhi<10^{14}\cdunits$ include a relatively
large fraction of WHIM absorbers.  However, a significant fraction of
absorbers in this range are still diffuse, and many WHIM or hot halo
absorbers have lower $b$-parameters than this threshold.

In summary, the linewidths trace temperature only fairly loosely, and
for any individual absorber it is difficult to pinpoint the cosmic
phase purely from observables.  Nevertheless, one can assign phases
probabilistically based on their location $(\nhi,b)$, at the penalty
of introducing some model dependence.  With the large ensembles of
absorbers expected to be obtained with COS, this approach can provide
some constraints on the content of all baryonic phases as traced by
\lya\ absorption.

\section{Conclusions}\label{sec:summary}

We study the low-redshift IGM as traced by \lya\ absorbers in $48\hmpc$,
$2\times 384^3$-particle cosmological hydrodynamic simulations with three
prescriptions for galactic outflows.  This work builds on our previous
works of a decade ago, namely \citet{dav99} and \citet{dav01}, by employing
simulations with $216\times$ more particles and $80\times$ more volume,
using improved numerical methodology, and most significantly, testing
the impact of large-scale galactic outflows on the IGM as traced by 
\lya\ absorption.  Our 
scenario for the nature and evolution of \lya\ forest absorbers from
$z=2\rightarrow 0$ remains basically unchanged: most absorbers arise
in low-overdensity, highly photoionized gas that traces uncollapsed
large-scale filaments.  \citet{dav99} argued that much of the observed
evolution in the \lya\ forest can be understood in terms of the evolving
relationship between \ion{H}{i} column density and the overdensity of
the absorbing gas; our latest simulations yield a similar slope and evolution
of this relation.  Outflows are found to have a small impact on the
diffuse IGM, but they do have a significant effect on bound-phase absorbers that
arise nearer to galaxies.
Our statistical results for \lya\ forest absorbers are obtained
by applying the AutoVP line-fitter \citep{dav97} to artificial spectra
with a resolution and S/N similar to those expected for typical
COS observations.

Our main results are as follows:
\begin{itemize}
\item Baryons at $z\sim 0$ are divided comparably amongst the 
diffuse phase, the WHIM, and the bound phases (condensed, hot halo, and stars),
The WHIM here is defined as low-density
(outside of halos) gas with $T>10^{5}$~K, a notable change from previous
definitions that impose a maximum temperature but not a maximum density.
In detail, our favoured momentum-driven outflow model has 41\% in diffuse
gas, 35\% in bound phases, and 24\% in WHIM at $z=0$, with values for
other wind models and redshifts shown in Figure~\ref{fig:baryphase}.
\item Our favoured momentum-driven wind (vzw) model shows similar
evolution in diffuse (unbound) IGM baryonic phases to the no-wind
(nw) case.  In contrast, the condensed, hot halo, and stellar fractions
are all noticeably different between those two, and between either
of these models and the
constant wind (cw) model.  Furthermore, cw significantly heats the IGM,
resulting in more WHIM and less condensed and diffuse gas.
\item The vzw model at COS resolution shows a 
column density distribution with a power-law slope of $-1.70$.
The slope is slightly steeper for nw and cw,
and it is steeper for higher-resolution spectra.  While formally all
slopes are within $\sim 2\sigma$ of each other, recall that the lines 
of sight are identical through each simulation, so the differences
are fairly robust.
\item The shallower vzw slope arises because it produces more strong
($\nhi>10^{14}\cdunits$) lines at $z=0$, which agrees better with
observations of $dN/dz$ evolution.  Weaker lines show no differences
in $dN/dz$ among wind models, but they 
do indicate a discrepancy around $z\sim 1$
that may reflect deviations from the \citet{haa01} ionizing background
evolution.
\item The linewidth ($b$) distribution shows a median around $\sim 40\kms$
at COS resolution, and it is insensitive to the wind model.  It is highly
sensitive to spectral resolution, even at large $b$ values, and much
of the broadening owes to instrumental resolution (note that our $b$-parameters
are not corrected for this).
\item Diffuse \lya\ absorbers follow a relation between $\nhi$ and
overdensity given by $\rho/\bar\rho \propto \nhi^{0.74}10^{-0.37z}$,
in general agreement with our previous findings.  This results in a
region of a given overdensity giving rise to a lower column density systems 
at higher redshift.  Departures from this relation become more prominent
at low-$z$ owing to an increased amount of shock heating.
\item The bulk of \lya\ absorbers have total hydrogen columns of
$\sim 10^{18-19}\cdunits$, reflecting the trend that the low-$\nhi$ absorbers
are more ionized (neutral fraction $\sim 10^{-6}$~cm$^{-3}$) and larger (hundreds
of kpc), while high-$\nhi$ absorbers are less ionized (neutral fraction
$\sim 10^{-3}$~cm$^{-3}$) and smaller (few kpc).
\item Linewidths only loosely reflect the temperature of the absorbing
gas, because typically linewidths are $1.5-3$ times the thermal
broadening value.  Hence, wide lines do not exclusively track WHIM
or hot gas, though trends in $\rho$ and $T$ with both $b$ and
$\nhi$ could be used to assign absorbers to a particular cosmic phase
in a statistical sense.
\item Improved $dN/dz$ statistics for weak lines ($\nhi < 10^{14}\cdunits$)
will improve determinations of the UV background intensity at $z<2$.
Improved statistics at higher column densities can provide strong tests of 
galactic outflow models.
\end{itemize}

The installation of COS on {\it Hubble} promises a great leap
forward in our understanding of low-redshift \lya\ absorption and the
IGM from which it arises.  Besides testing the basic paradigm of the
cosmic web origin for weak \lya\ forest absorbers, COS \lya\ data will
provide constraints on the baryon content in various cosmic phases,
ionizing background evolution, and the nature of galactic outflows.
Additional constraints will be provided by the distribution of IGM metals
that can be probed to much greater precision with COS (the subject of
Paper~II in this series) and by the relation between galaxies and
absorption systems (covered in Paper~III).  
The combination of these constraints with overall \lya\ absorber
statistics will provide powerful diagnostics for the mechanisms
of galactic outflows.
Comparisons of theory and observations 
in the coming years will critically test 
our prevailing paradigms and prescriptions,
thereby providing valuable insights into the evolution of the majority
of baryons over cosmic time.

 \section*{Acknowledgements}
We thank M. Fardal, D. Kere\v{s}, N. Lehner, M. Peeples, and 
J. Schaye for helpful
conversations, and G. Williger for providing us the data shown in
Figure~\ref{fig:dndz} in electronic form.  The simulations used here were
run on University of Arizona's SGI cluster, ice.  Support for this work
was provided by NASA through grant numbers HST-GO-11598 and HST-AR-11751
from the Space Telescope Science Institute, which is operated by AURA,
Inc. under NASA contract NAS5-26555.  This work was also supported
by the National Science Foundation under grant number AST-0847667.
Computing resources were obtained through grant number DMS-0619881 from
the National Science Foundation.  DW acknowledges support from NSF grant
AST-0707985 and from an AMIAS membership at the Institute for Advanced
Study.  NK acknowledges support from NASA through ADP grant NNX08AJ44G.

\end{document}